\documentclass[10pt, twocolumn, tighten]{aastex62}

\usepackage{color}
\usepackage{comment}
\usepackage{graphicx}

\usepackage{hyperref}
\usepackage{url}

\newcommand{\beit}{\begin{itemize}}
\newcommand{\enit}{\end{itemize}}

\newcommand{\hi}{H{\sc i}}
\newcommand{\lya}{Ly$\alpha$}

\newcommand{\hii}{H{\sc ii}}

\newcommand{\Civ}{{\sc Civ}}
\newcommand{\oii}{{\sc [Oii]}}

\newcommand{\sqdeg}{deg$^{2}$}
\newcommand{\hmpc}{$h^{-1}$ cMpc}

\newcommand{\hrp}{\hi\ radial profile}
\newcommand{\rthd}{$R_{\rm 3D}$}

\newcommand{\dF}{$\delta_{F}$}

\newcommand{\dQSO}{$\delta_{\rm QSO}$}

\newcommand{\beq}{\begin{equation}}
\newcommand{\eeq}{\end{equation}}

\newcommand{\dperp}{\ensuremath{\langle d_{\perp} \rangle}}

\newcommand{\delrecon}{\ensuremath{\delta_F^{\rm rec}}}
\newcommand{\delobson}{\ensuremath{\delta_F^{\rm obs}}}

\newcommand{\cmd}{\mathbf{C}_\mathrm{MD}}
\newcommand{\cdd}{\mathbf{C}_\mathrm{DD}}

\begin{document}

\title{
Cosmological 3D \hi\ Gas Map with HETDEX Ly$\alpha$ Emitters and eBOSS QSOs at $z=2$:\\
IGM-Galaxy/QSO Connection and a $\sim$ 40-Mpc Scale Giant \hii\ Bubble Candidate}
\shorttitle{\hi\ Gas Map with HETDEX Ly$\alpha$ Emitters and eBOSS QSOs}
\accepted{for Publication in ApJ}
\shortauthors{Mukae et al.}
 
\author{Shiro Mukae} 
\affiliation{
Institute for Cosmic Ray Research, The University of Tokyo, 
5-1-5 Kashiwanoha, Kashiwa, Chiba 277-8582, Japan}
\affiliation{
Department of Astronomy, Graduate School of Science, The University of Tokyo, 
7-3-1 Hongo, Bunkyo, Tokyo, 113-0033, Japan}
\email{mukae@icrr.u-tokyo.ac.jp}
\author{Masami Ouchi}
\affiliation{
National Astronomical Observatory of Japan, 2-21-1 Osawa, Mitaka, Tokyo 181-8588, Japan}
\affiliation{
Institute for Cosmic Ray Research, The University of Tokyo, 
5-1-5 Kashiwanoha, Kashiwa, Chiba 277-8582, Japan}
\affiliation{
Kavli Institute for the Physics and Mathematics of the Universe 
(Kavli IPMU, WPI), The University of Tokyo, 
5-1-5 Kashiwanoha, Kashiwa, Chiba, 277-8583, Japan}
\author{Gary J. Hill}
\affiliation{
Department of Astronomy, The University of Texas at Austin, 2515 Speedway, Stop C1400, Austin, Texas 78712, USA}
\affiliation{
McDonald Observatory, University of Texas at Austin, 2515 Speedway, Stop C1402, Austin, TX 78712, USA}
\author{Karl Gebhardt}
\affiliation{
Department of Astronomy, The University of Texas at Austin, 2515 Speedway, Stop C1400, Austin, Texas 78712, USA}
\author{Erin Mentuch Cooper}
\affiliation{
Department of Astronomy, The University of Texas at Austin, 2515 Speedway, Stop C1400, Austin, Texas 78712, USA}
\author{Donghui Jeong}  
\affiliation{Department of Astronomy and Astrophysics, 
The Pennsylvania State University, University Park, PA 16802} \affiliation{Institute for Gravitation and the Cosmos, The Pennsylvania State University, University Park, PA 16802}
\author{Shun Saito}  
\affiliation{
Institute for Multi-messenger Astrophysics and Cosmology, Department of Physics, 
Missouri University of Science and Technology, 1315 N. Pine Street, Rolla, MO 65409, USA}
\affiliation{
Kavli Institute for the Physics and Mathematics of the Universe 
(Kavli IPMU, WPI), The University of Tokyo, 
5-1-5 Kashiwanoha, Kashiwa, Chiba, 277-8583, Japan}
\author{Maximilian Fabricius}  
\affiliation{Max-Planck-Institut f\"{u}r extraterrestrische Physik, Giessenbachstrasse, D-85748 Garching, Germany}
\affiliation{Universit\"{a}ts-Sternwarte M\"{u}nchen, Scheinerstrasse 1, D-81679 M\"{u}nchen, Germany}
\author{Eric Gawiser}
\affiliation{Department of Physics and Astronomy, Rutgers University,
Piscataway, NJ 08854, USA}
\author{Robin Ciardullo}  
\affiliation{Department of Astronomy and Astrophysics, 
The Pennsylvania State University, University Park, PA 16802} \affiliation{Institute for Gravitation and the Cosmos, The Pennsylvania State University, University Park, PA 16802}
\author{Daniel Farrow}  
\affiliation{Max-Planck-Institut f\"{u}r extraterrestrische Physik, Giessenbachstrasse, D-85748 Garching, Germany}
\author{Dustin Davis}
\affiliation{
Department of Astronomy, The University of Texas at Austin, 2515 Speedway, Stop C1400, Austin, Texas 78712, USA}
\author{Greg Zeimann} 
\affiliation{
Hobby Eberly Telescope, University of Texas, Austin, TX 78712, USA}
\author{Steven L. Finkelstein}   
\affiliation{
Department of Astronomy, The University of Texas at Austin, 2515 Speedway, Stop C1400, Austin, Texas 78712, USA} 
\author{Caryl Gronwall}
\affiliation{Department of Astronomy and Astrophysics, 
The Pennsylvania State University, University Park, PA 16802}
\affiliation{Institute for Gravitation and the Cosmos, The Pennsylvania State University, University Park, PA 16802}
\author{Chenxu Liu}   
\affiliation{
Department of Astronomy, The University of Texas at Austin, 2515 Speedway, Stop C1400, Austin, Texas 78712, USA}
\author{Yechi Zhang}  
\affiliation{
Institute for Cosmic Ray Research, The University of Tokyo, 
5-1-5 Kashiwanoha, Kashiwa, Chiba 277-8582, Japan}
\affiliation{
Department of Astronomy, Graduate School of Science, The University of Tokyo,
7-3-1 Hongo, Bunkyo, Tokyo, 113-0033, Japan}
\author{Chris Byrohl}  
\affiliation{
Max-Planck-Institut f\"{u}r Astrophysik, Karl-Schwarzschild-Str. 1, D-85748 Garching, Germany}
\author{Yoshiaki Ono} 
\affiliation{
Institute for Cosmic Ray Research, The University of Tokyo, 
5-1-5 Kashiwanoha, Kashiwa, Chiba 277-8582, Japan}
\author{Donald P. Schneider}
\affiliation{Department of Astronomy and Astrophysics, 
The Pennsylvania State University, University Park, PA 16802}
\affiliation{Institute for Gravitation and the Cosmos, The Pennsylvania State University, University Park, PA 16802}
\author{Matt J. Jarvis } 
\affiliation{Astrophysics, Department of Physics, Keble Road, Oxford, OX1 3RH, UK}
\affiliation{Department of Physics \& Astronomy, University of the WesternCape, Private Bag X17, Bellville, Cape Town, 7535, South Africa}
\author{Caitlin M. Casey}   
\affiliation{
Department of Astronomy, The University of Texas at Austin, 2515 Speedway, Stop C1400, Austin, Texas 78712, USA}
\author{Ken Mawatari}   
\affiliation{
Institute for Cosmic Ray Research, The University of Tokyo, 
5-1-5 Kashiwanoha, Kashiwa, Chiba 277-8582, Japan}

\begin{abstract}
We present cosmological ($30-400$ Mpc) distributions of neutral hydrogen (\hi) in the inter-galactic medium (IGM) traced by \lya\ Emitters (LAEs) and QSOs at $z=2.1-2.5$, 
selected with the data of the on-going Hobby-Eberly Telescope Dark Energy Experiment (HETDEX) and the eBOSS survey. 
Motivated by a previous study of \citet{Mukae2020a}, 
we investigate spatial correlations of LAEs and QSOs 
with \hi\ tomography maps reconstructed from 
\hi\ \lya\ forest absorption in the spectra of background galaxies and QSOs 
obtained by the CLAMATO survey and this study, respectively. 
In the cosmological volume 
far from QSOs, we find that LAEs reside in regions of strong \hi\ absorption, i.e. \hi\ rich, which is consistent with results of previous galaxy-background QSO pair studies.
Moreover, there is an anisotropy in the \hi-distribution plot of transverse and line-of-sight distances; on average the \hi\ absorption peak is blueshifted by $\sim 200$ km s$^{-1}$ from the LAE \lya\ redshift, reproducing the known average velocity offset between the \lya\ emission redshift and the galaxy systemic redshift.
We have identified a $\sim$ 40-Mpc scale volume of \hi\ underdensity 
that is a candidate for a giant \hii\ bubble, where six QSOs and 
an LAE overdensity exist at $\left < z \right > =2.16$. 
The coincidence of the QSO and LAE overdensities with the \hi\ underdensity indicates that 
the ionizing photon radiation of the QSOs has created a highly ionized volume of multiple proximity zones in a matter overdensity. Our results suggest an evolutionary picture where \hi\ gas in an overdensity of galaxies becomes highly photoionized when QSOs emerge in the galaxies.
\end{abstract}

\keywords{
galaxies: formation --- 
intergalactic medium  --- 
large-scale structure of universe
}

\section{Introduction} \label{sec:introduction}
In the modern paradigm of galaxy formation, 
galaxies form and evolve in gaseous filamentary structures 
\citep[e.g.,][]{Mo2010a, Meiksin2009a}.
Studies of cosmological hydrodynamics simulations have suggested 
a picture where galaxies and the gaseous large-scale structures (LSSs)
exchange baryonic gas by gas flows \citep{Fox2017a, Voort2017a}.
Cold gas ($\sim10^{4} \rm\ K$) in the intergalactic medium (IGM) 
accretes onto galaxies through the filamentary structures, 
and triggers star formation in the galaxies 
\citep[e.g.,][]{Dekel2009a, Keres2005a}.
Star formation heats up the gas, and the gas is expelled from the galaxies 
by feedback processes such as galactic outflows
\citep[e.g.,][]{Somerville2015a, Viel2013a}. 
Observing the site of the gas exchange is key for understanding 
galaxy formation in gaseous LSSs, especially at $z\sim2-3$ 
when the star-formation rate density peaks in cosmic history.
However, the connections between galaxies and gaseous LSSs 
are as yet poorly probed in observations.
\par
To study galaxy formation in gaseous LSSs, 
recent observational studies have probed 
the IGM neutral hydrogen (\hi) at $z\sim2-3$, 
revealing the spatial distribution of 
the \hi\ \lya\ forest absorption (\hi\ absorption). 
Until a decade ago, \hi-gas distributions around galaxies 
were studied by stacking analyses of background QSO spectra 
in which the \hi-gas of foreground galaxies causes weak absorption \citep[e.g.,][]{Adelberger2005a, Turner2014a, Bielby2017a}.
The measurements of stacked spectra have shown \hi\ absorption 
as a function of transverse distance to the background QSO sightline, 
and revealed an \hi\ absorption excess around massive star-forming galaxies 
over $\sim 5$ $h^{-1}$ comoving Mpc ($h^{-1}$ cMpc). 
However, the stacked \hi-gas distributions
are based on measurements in multiple fields,
and represent the cosmic-averaged distribution, 
losing information on specific galaxy environments 
such as overdensities of galaxies and QSOs.
\par
In the past few years, 
\cite{Lee2014b} have observationally demonstrated a
\hi\ tomography technique that reconstructs 
three dimensional (3D) \hi\ LSSs at $z\sim2$ 
from \hi\ absorption found in multiple background galaxy spectra. 
The \hi\ tomography technique was originally proposed 
by \cite{Pichon2001a} and \cite{Caucci2008a} 
for the purpose of recovering the large-scale topology of 
the underlying matter field.
The observational requirements of \hi\ tomography are 
investigated by \cite{Lee2014a} for 8--10m-class telescopes and by
\citealt{Steidel2009a} and \citealt{Evans2012a},
for those of 30 m-class.
The subsequent \hi\ tomography studies of \cite{Lee2016a, Lee2018a} have revealed \hi\ LSSs with spatial resolutions of $2.5$ \hmpc\ 
in the COSMOS \lya\ Mapping And Tomography Observations (CLAMATO) survey.
Although the \hi\ tomography technique has enabled 
spatial characterization of \hi\ LSSs in a field of interest, 
only a few studies systematically investigate 
connections between \hi\ LSSs and galaxies
in a range of environments 
from blank fields
(that are randomly selected extra-galactic survey fields)
to specific fields such as galaxy overdensities \citep{Mukae2020a, 
Newman2020a}.
\par 
As a wide-field and statistical study complementary to the CLAMATO survey, \cite{Mukae2017a} have investigated spatial correlations of 
average \hi-gas overdensities and galaxy overdensities at $z\sim 2-3$, using galaxy photometry and multiple spectra of background QSOs in a large 1.62 \sqdeg\ area of the COSMOS/UltraVISTA field.
The spatial correlation results suggest that 
a large amount of \hi-gas is associated with galaxy overdensities
(see also \citealt{Liang2020a, Nagamine2020a}).
However, it is still unknown how the distribution of the \hi\ LSSs is affected 
by overdensities of galaxies and QSOs on an individual structure basis. 
Because QSOs often emerge in galaxy overdensities, the QSOs' radiation can enhance the ultraviolet background (UVB) radiation in the overdensities, photo-ionizing the surrounding \hi-gas \citep[e.g.,][]{Umehata2019a, Kikuta2019}. 
These QSO photoionization regions are dubbed proximity zones
whose sizes are observationally estimated to be 
$\sim 10-15\ h^{-1}$ cMpc in diameter at $z\sim2$ 
\citep[e.g.,][]{DOdorico2008a, Mukae2020a, Jalan2019a}.
Moreover, enhanced UVB radiation can suppress star formation of galaxies 
in low-mass dark matter halos by photo-evaporation of their gas \cite[e.g.,][]{Susa2004a, Susa2000a}, which is implied by observational studies of galaxy number counts around QSOs \citep[e.g.,][]{Kashikawa2007a, Kikuta2017a}. 
The three key elements for galaxy formation in LSSs 
are dark matter, gas, and ionization.
To understand the impact of overdensities of galaxies and QSOs on 
the surrounding gas, systematic study 
the \hi-gas distributions around galaxies in various galaxy environments is required.
\par
In this study, we investigate IGM \hi-gas distributions around galaxies at $z\sim2$ 
in the following two galaxy environments: 
a blank field (i.e. a randomly selected extra-galactic survey field) and 
an extreme QSO overdensity region.
We use the large datasets of galaxies and QSOs consisting of the spectroscopic data of the Hobby-Eberly Telescope Dark Energy Experiment 
(HETDEX; \citealt{Hill2008a, Hill2016a}, Gebhardt et al. 2020, in preparation)
and the extended Baryon Oscillation Spectroscopic Survey of 
the Sloan Digital Sky Survey IV \citep[SDSS-IV/eBOSS;][]{Dawson2016a}, respectively. 
To probe IGM \hi-gas distributions at $z\sim2$, 
we use \hi\ absorption found in spectra of background QSOs at $z>2$.
We perform \hi\ tomography based on the multiple \hi\ absorptions,
to reveal 3D \hi\ LSSs around the galaxies.
\par 
The structure of this paper is as follows. 
In Section \ref{sec: galaxy and quasar catalogs},
we describe our datasets of foreground/background galaxies and QSOs.
In Section \ref{sec: tom tech and map}, 
we detail our \hi\ tomography techniques and our \hi\ tomography maps.
We present results and discussion 
in Sections \ref{sec: results} and \ref{sec: discussion}, respectively.
In Section \ref{sec: summary}.
we summarize our major findings. 
Throughout this paper, we use a cosmological parameter set:
$( \Omega_m, \Omega_\Lambda, \Omega_b, h)$=$( 0.26, 0.74, 0.045, 0.70)$ 
consistent with the nine-year $WMAP$ result \citep{Hinshaw2013a}.
We refer to kpc and Mpc in comoving (physical) 
units as ckpc and cMpc (pkpc and pMpc), respectively.
We specifically use units including $h^{-1}$ for ckpc and cMpc, because these units are widely found in this field of study, and allow readers to easily compare our results with previous ones.
All magnitudes are in AB magnitudes \citep{Oke1983a}.

\section{Galaxy and QSO Catalogs} \label{sec: galaxy and quasar catalogs}
We investigate IGM \hi-gas distributions around $z\sim2$ galaxies  
in two fields shown in Sections \ref{sec: results cosmos} and \ref{sec: results egs},
respectively:
\beit
\item COSMOS: a blank field of $0.157$ \sqdeg\ 
with no extreme overdensities
that is placed around the center of 
the CLAMATO survey \citep{Lee2018a} 
\item EGS: a field of $6.0$ \sqdeg\ 
that is a combination of 
the original and flanking EGS \citep{Davis2007} regions.
Although this is a blank field (i.e. a randomly selected extra-galactic survey field), it contains a significant QSO overdensity (Section \ref{sec: results egs}). 
\enit 

\par 
We describe our galaxy (i.e. LAE) and QSO catalogs 
in Sections \ref{sec: lae_catalog} and \ref{sec: quasar_catalog}, respectively. 
The QSO catalogs contain foreground and background QSOs 
(Sections \ref{sec: foreground_qusasar}-\ref{sec: background_qusasar}).
The background QSOs are used for probing \hi\ absorption 
via an \hi\ tomography map covering $z=2.05-2.55$, 
while the foreground QSOs are those included in 
a cosmic volume of the \hi\ tomography map. 
Note that, throughout this paper, 
we use the words 'foreground' and 'background' 
for sources located in the \hi\ tomography maps and 
for those utilized to create the \hi\ tomography map, respectively.
\footnote{
Although the background sources reside at $z=2.1-3.1$, 
the redshift range of \lya\ forest absorption 
contributing to the \hi\ tomography map construction 
depends on the background-source redshift. 
This is because \lya\ forest absorption over rest-frame wavelengths $1041-1185$\AA\ 
is used for the \hi\ tomography mapping.
For this reason, we do not associate a particular redshift range with the background sources, but use the term 'background' throughout the paper.}
Table \ref{tab:data} summarizes the data sources for the galaxies and QSOs.

\begin{deluxetable*}{cccccc}[ht]
\tablecolumns{6} 
\tablewidth{0pt}
\vspace{-0.5em}
\tablecaption{ Data Summary of the Two Fields \label{tab:data} }
\tablehead{
\colhead{Field} &  \colhead{Area} &  \colhead{Volume} &  \colhead{Galaxy} &  \colhead{QSO}  & \colhead{\hi\ Map}\\ 
\colhead{ }   & \colhead{(deg$^2$)} & \colhead{($h^{-3}$ cMpc$^3$)} & \colhead{Sample} &  \colhead{Sample} & \colhead{}}
\startdata 
COSMOS & 0.157 & $3.2\times10^5$ & HETDEX & eBOSS & \cite{Lee2018a} \\
EGS & 6.0 & $7.5\times10^6$ $^\dagger$ & HETDEX & eBOSS & This Study 
\enddata 
\centering
\tablenotetext{^{\dagger}}{
In EGS, the QSO overdensity EGS-QO1 (Section \ref{sec: results egs}) occupies a volume of $1.1\times10^5$ $h^{-3}$ cMpc$^3$.}
\end{deluxetable*}

\subsection{LAE Catalogs} \label{sec: lae_catalog}
Two catalogs of foreground galaxies are drawn from samples of \lya\ emitters (LAEs) in the COSMOS and EGS regions.
The COSMOS and EGS LAE catalogs are constructed 
from early observations of HETDEX, 
obtained with the Visible Integral-field Replicable Unit Spectrograph (VIRUS, \citealt{hill2018a}) 
on the upgraded 10 m Hobby-Eberly Telescope (HET).
The HET \citep{Ramsey1994a, Hill2018b} is an innovative telescope 
with 11-meter segmented primary mirror, 
located in West Texas at the McDonald Observatory.
VIRUS is a massively replicated 
integral field spectrograph \citep{Hill2014a},
designed for blind spectroscopic surveys. 
It consists of 78 fiber integral field units (IFUs; \citealt{Kelz2014a})
distributed within the 22 arcmin diameter field of view of the telescope. 
A detailed technical description of the HET wide field upgrade and VIRUS 
is presented in Hill et al. (2020, in preparation).\footnote{
The VIRUS array has been undergoing staged deployment of 
IFUs and spectrograph units starting in late 2015. 
The data presented in this paper were obtained with between 16 and 47 active IFUs, with up to 21,056 fibers
and were observed between January 3, 2017 to February 09, 2019.}
Each IFU feeds 448  fibers with diameter $1 \farcs 5$ to a pair of spectrographs.
The spectrographs have a fixed spectral bandpass of
$\lambda =3500$--$5500$ \AA\ and a spectral resolution of $R \approx 800$.
The HETDEX program is performing a blind emission line survey 
over a total of $\sim 450$ \sqdeg\ area 
with a standard exposure set of 6 min $\times$ 3 dithers 
(to fill the sky gaps between fibers), and 
aims to identify 10$^6$ LAEs at \lya-emission redshifts of
$z_{\rm Ly\alpha}=1.9$--$3.5$ in a $9$ Gpc$^3$ volume.
The HETDEX survey constructs an emission-line database 
(Gebhardt et al. 2020, in preparation)
where emission-line detections are processed in combination
with broadband imaging data, including data from Subaru/Hyper Suprime-Cam (HSC) 
(Gebhardt et al. 2020 in preparation). 
\par
\begin{figure*}[p]
\begin{tabular}{c}
\begin{minipage}[t]{1.0\hsize}
\centering
 \includegraphics[width=0.6\hsize, clip, bb=0 0 900 775, clip=true]{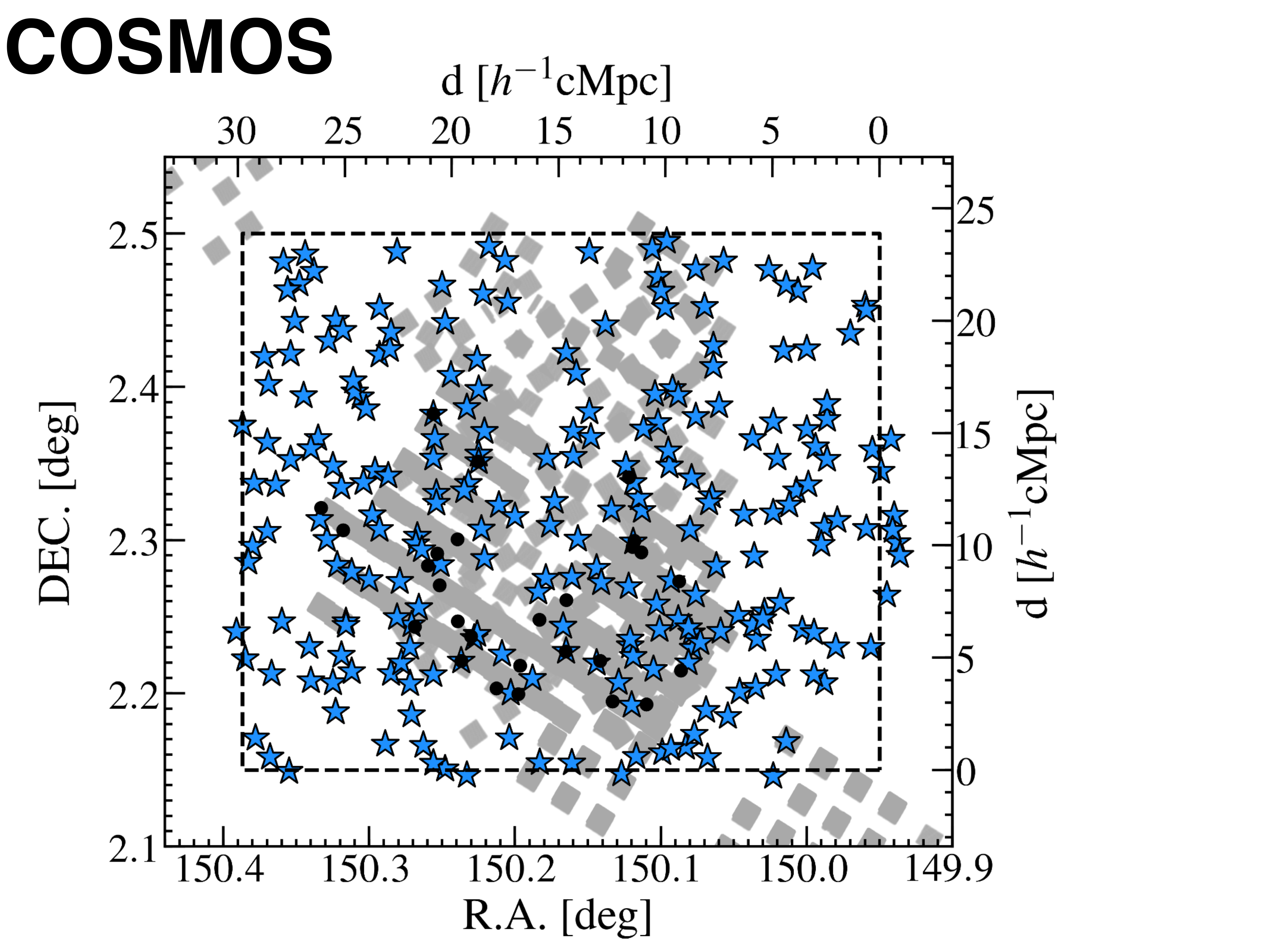}
\caption{
 Sky distribution of galaxies and QSOs
 in the $0.157$ \sqdeg\ area of the COSMOS field.  
 The blue stars represent the positions of background objects 
 that are used for \hi\ tomography mapping \citep{Lee2018a}. 
 The black dots are the foreground galaxies, HETDEX LAEs, at $z=2.05$--$2.55$ (Section \ref{sec: lae_catalog}). 
 Note that no foreground QSOs are found 
 in this moderately small area of the COSMOS field.
 The gray shaded squares present the area covered by the IFUs of the HETDEX survey.
 The dashed-line box indicates the COSMOS field for the \hi\ tomography map 
 whose spatial resolution is 2.5 \hmpc\  (Section \ref{subsec: tom cosmos}).
 The axes on the top and right-hand side indicate the projected comoving scale at $z=2.3$.}
 \label{fig: sky cosmos}
\end{minipage}\\
\begin{minipage}[t]{1.0\hsize}
\centering
 \includegraphics[width=0.6\hsize, clip, bb=0 0 900 775, clip=true]{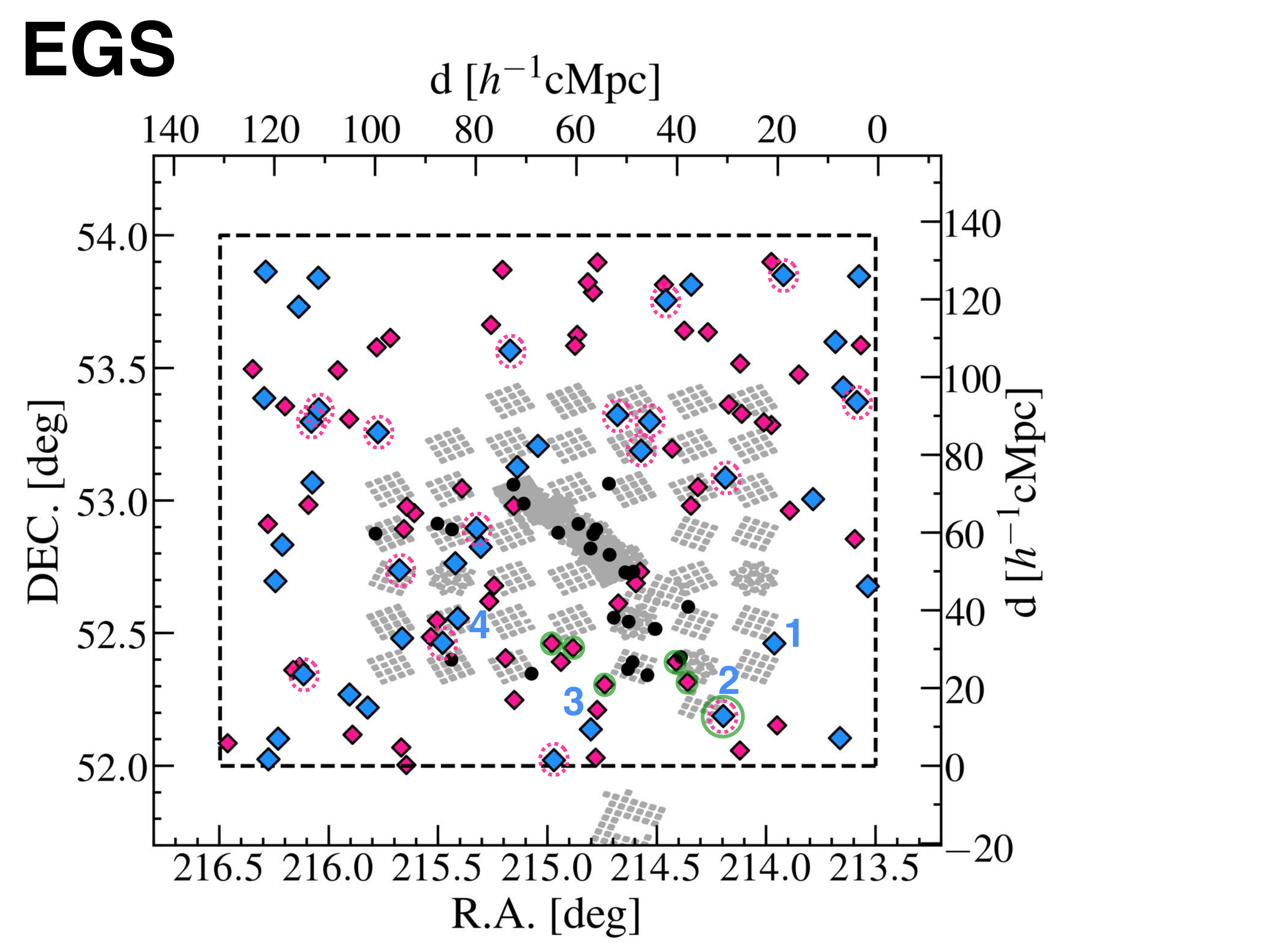}
 \caption{
 Sky distribution of foreground and background QSOs 
 in the $6.0$ \sqdeg\ area of the EGS field. 
 The blue diamonds are the positions of the background QSOs 
 (Section \ref{sec: background_qusasar}). 
 The background QSO sightlines 1-4 probe a large \hi\ underdensity 
 (Section \ref{sec: results egs}) and their spectra are presented in Figure \ref{fig: bq spec 2}.
 The magenta diamonds are the positions of foreground QSOs at $z=2.05$--$2.55$ (Section \ref{sec: foreground_qusasar}).  
 The magenta dotted circles indicate the background QSOs 
 that are also used as foreground QSOs (Section \ref{sec: results egs}).
 The green circles represent six QSOs comprising the extreme QSO overdensity, EGS-QO1 (Section \ref{sec: results egs}).
 The black dots are HETDEX LAEs at $z=2.05$--$2.55$ 
 (Section \ref{sec: lae_catalog}).
 The gray shaded squares present the area covered by the IFUs of the HETDEX survey.
 The dashed-line box is the EGS field for the \hi\ tomography map 
 whose spatial resolution is 20 \hmpc\  (Section \ref{subsec: tom egs}).
 The axes on the top and right-hand side indicate 
 the projected comoving scale at $z=2.3$.
 }
 \label{fig: sky egs}
\end{minipage}
\end{tabular}
\end{figure*}

\begin{figure*}[p]
\begin{tabular}{c}
\begin{minipage}[t]{1,0\hsize}
\centering
\includegraphics[width=1.0\hsize, clip, bb= 0 0 1100 500, clip=true]{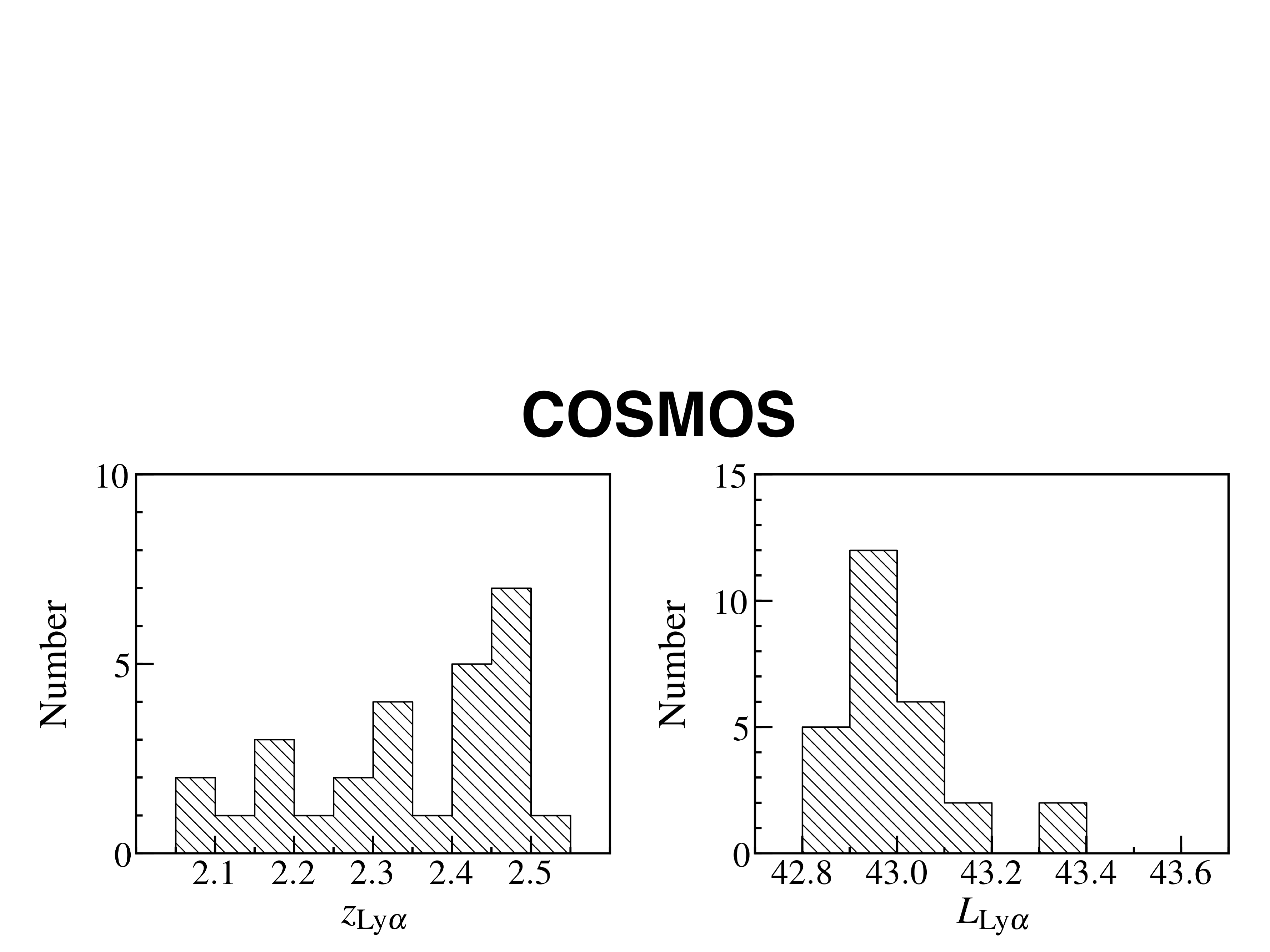}
\caption{
 $z_{\rm Ly\alpha}$ (left) and $L_{\rm Ly\alpha}$ (right)
 distributions of 27 HETDEX LAEs
 in the $0.157$ \sqdeg\ area of the COSMOS field.  
 }
\label{fig: hist_lae_cosmos}
\end{minipage}\\
\begin{minipage}[t]{1,0\hsize}
\centering
\includegraphics[width=1.0\hsize, clip, bb= 0 0 1100 500, clip=true]{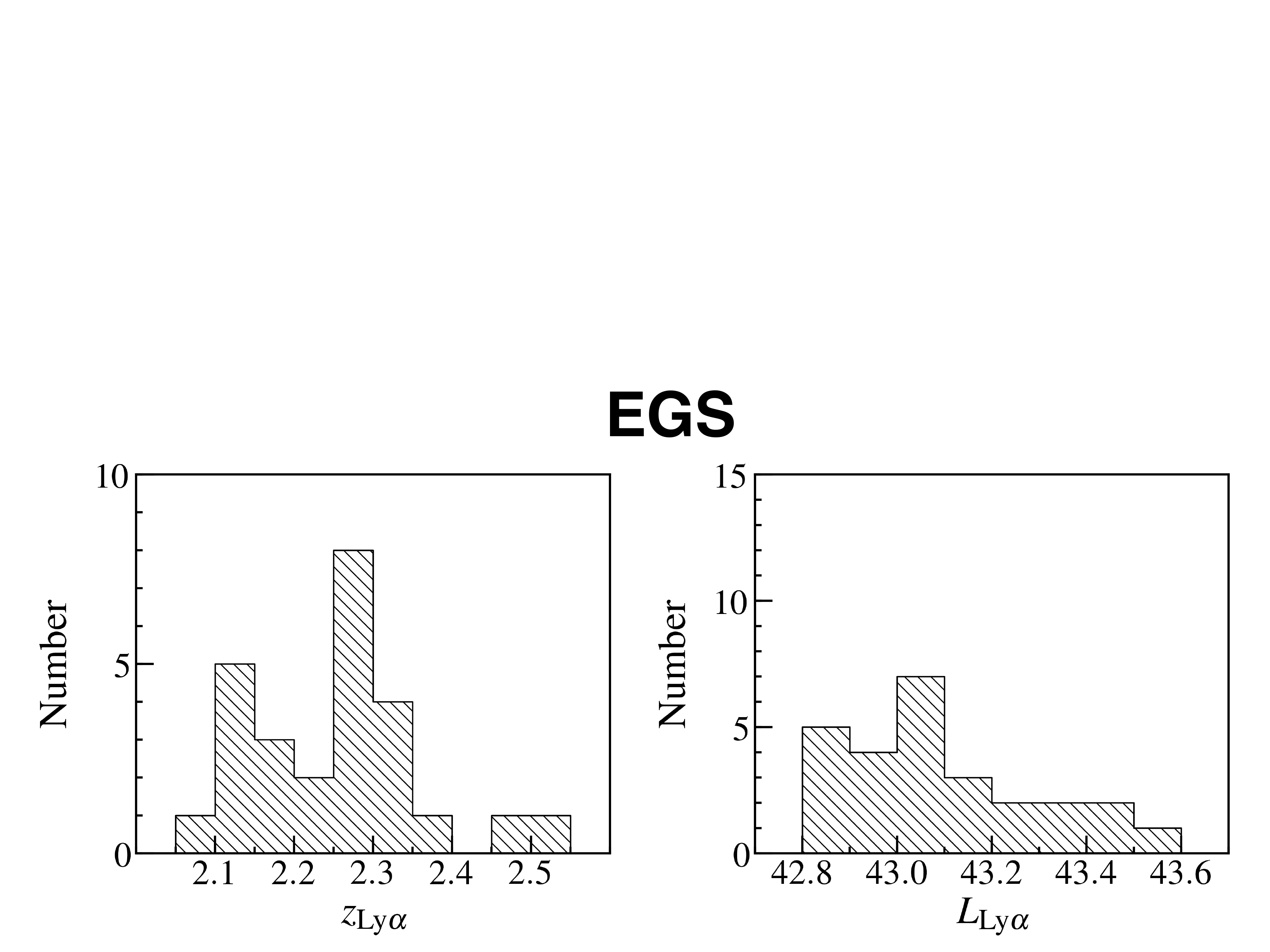}
\caption{
 $z_{\rm Ly\alpha}$ (left) and $L_{\rm Ly\alpha}$ (right)
 distributions of 26 HETDEX LAEs
 in the $6.0$ \sqdeg\ area of the EGS field. 
 }
\label{fig: hist_lae_egs}
\end{minipage}
\end{tabular}
\end{figure*}

 
We choose 27 (26) 
spectroscopically identified LAEs in the COSMOS (EGS) field 
by the following three criteria:
i) a single emission-line is detected with a significance level greater than 6.5 $\sigma$ as defined by the HETDEX line-identification algorithm (Gebhardt et al. 2020, in preparation) 
that considers the fiber filling factor within the IFUs of $1/3$, 
ii) a high observed equivalent width of the single emission line and 
a low luminosity, 
distinguishing high-$z$ Ly$\alpha$ from low-$z$ \oii\ emission
with Bayesian statistics 
whose prior distributions are given by previous optical spectroscopic results with a wide-wavelength coverage \citep{Leung2017},
iii) the total luminosity of the \lya\ emission is $L_{\rm Ly\alpha} > 10^{42.8}$ erg s$^{-1}$, which achieves source identification completeness of $\gtrsim 90$\% (Gebhardt et al. 2020, in preparation; Zhang et al. in preparation), 
and
iv) the redshift of the \lya\ emission line falls in the range of $z_{\rm Ly\alpha}=2.05$--$2.55$.
The redshift range is chosen to match with that of the COSMOS \hi\ tomography map \citep[][Section \ref{subsec: tom cosmos}]{Lee2018a}.
Figures \ref{fig: sky cosmos} and \ref{fig: sky egs} 
(Figures \ref{fig: hist_lae_cosmos} and \ref{fig: hist_lae_egs}) 
present the sky (redshift and luminosity) distributions of 
our HETDEX LAEs in the COSMOS and EGS fields, respectively.
The basic properties of the HETDEX LAEs in the COSMOS and EGS fields 
are summarized in Tables \ref{tab: fg cosmos} and \ref{tab: fg egs}, respectively.

\begin{deluxetable*}{ccccc}[ht]
\tablecolumns{5} 
\tablewidth{0pt}
\vspace{-0.5em}
\tablecaption{ HETDEX LAEs in the COSMOS field \label{tab: fg cosmos} }
\tablehead{
\colhead{ID} &  \colhead{R.A.} &  \colhead{Decl.} &  \colhead{$z_{\rm Ly\alpha}$} &  \colhead{$L_{\rm Ly\alpha}$}  \\ 
\colhead{ }   & \colhead{(J2000)} & \colhead{(J2000)} & \colhead{} &  \colhead{($10^{42}$ erg s$^{-1}$)} }
\startdata 
HETDEX J100028.34+021758.5 & 10:00:28.34 & +02:17:58.50 & 2.492 & 9.28 \\
HETDEX J100051.07+021211.9 & 10:00:51.07 & +02:12:11.94 & 2.444 & 11.42 \\
HETDEX J100043.97+021452.8 & 10:00:43.97 & +02:14:52.76 & 2.486 & 24.69 \\
HETDEX J100020.68+021253.3 & 10:00:20.68 & +02:12:53.32 & 2.156 & 8.47 \\
HETDEX J100116.29+021823.0 & 10:01:16.29 & +02:18:23.04 & 2.322 & 7.56 \\
HETDEX J100031.95+021140.8 & 10:00:31.95 & +02:11:40.80 & 2.198 & 8.72 \\
HETDEX J100056.83+021316.3 & 10:00:56.83 & +02:13:16.32 & 2.433 & 13.23 \\
HETDEX J100102.35+021659.9 & 10:01:02.35 & +02:16:59.89 & 2.508 & 19.97 \\
HETDEX J100054.09+022104.8 & 10:00:54.09 & +02:21:04.80 & 2.472 & 9.34 \\
HETDEX J100104.47+021436.5 & 10:01:04.47 & +02:14:36.50 & 2.139 & 6.58 \\
HETDEX J100057.47+021801.7 & 10:00:57.47 & +02:18:01.68 & 2.163 & 10.05 \\
HETDEX J100021.01+021622.9 & 10:00:21.01 & +02:16:22.92 & 2.441 & 8.35 \\
HETDEX J100119.92+021915.8 & 10:01:19.92 & +02:19:15.82 & 2.323 & 9.43 \\
HETDEX J100100.42+021613.9 & 10:01:00.42 & +02:16:13.86 & 2.099 & 7.23 \\
HETDEX J100047.46+021158.1 & 10:00:47.46 & +02:11:58.05 & 2.282 & 8.51 \\
HETDEX J100039.54+021539.0 & 10:00:39.54 & +02:15:38.96 & 2.453 & 10.89 \\
HETDEX J100028.65+021744.0 & 10:00:28.65 & +02:17:44.05 & 2.099 & 9.47 \\
HETDEX J100101.45+022256.5 & 10:01:01.45 & +02:22:56.51 & 2.320 & 6.34 \\
HETDEX J100047.17+021305.1 & 10:00:47.17 & +02:13:05.06 & 2.340 & 9.58 \\
HETDEX J100027.22+021731.5 & 10:00:27.22 & +02:17:31.47 & 2.287 & 11.31 \\
HETDEX J100100.82+021728.7 & 10:01:00.82 & +02:17:28.67 & 2.470 & 9.31 \\
HETDEX J100026.37+021134.2 & 10:00:26.37 & +02:11:34.22 & 2.376 & 11.97 \\
HETDEX J100029.24+022027.3 & 10:00:29.24 & +02:20:27.27 & 2.467 & 13.17 \\
HETDEX J100057.43+021449.5 & 10:00:57.43 & +02:14:49.48 & 2.499 & 7.41 \\
HETDEX J100033.97+021316.2 & 10:00:33.97 & +02:13:16.15 & 2.230 & 9.26 \\
HETDEX J100055.21+021413.7 & 10:00:55.21 & +02:14:13.67 & 2.414 & 9.21 \\
HETDEX J100039.63+021338.3 & 10:00:39.63 & +02:13:38.35 & 2.441 & 11.37
\enddata 
\centering
\end{deluxetable*}

\begin{deluxetable*}{cccccc}[ht]
\tablecolumns{6} 
\tablewidth{0pt}
\vspace{-0.5em}
\tablecaption{ HETDEX LAEs in the EGS field \label{tab: fg egs} }
\tablehead{
\colhead{ID} &  \colhead{R.A.} &  \colhead{Decl.} &  \colhead{$z_{\rm Ly\alpha}$} &  \colhead{$L_{\rm Ly\alpha}$} & \colhead{Label\tablenotemark{a}}  \\ 
\colhead{ }   & \colhead{(J2000)} & \colhead{(J2000)} & \colhead{} &  \colhead{($10^{42}$ erg s$^{-1}$)} & \colhead{}  }
\startdata 
HETDEX J141948.50+525246.9 & 14:19:48.50 & +52:52:46.85 & 2.206 & 9.85 & - \\
HETDEX J141909.77+525223.4 & 14:19:09.77 & +52:52:23.38 & 2.296 & 11.98 & - \\
HETDEX J141851.86+524745.9 & 14:18:51.86 & +52:47:45.85 & 2.451 & 7.58 & - \\
HETDEX J141913.02+524911.2 & 14:19:13.02 & +52:49:11.23 & 2.156 & 11.38 & LAE4 \\
HETDEX J141906.49+525328.0 & 14:19:06.49 & +52:53:27.98 & 2.531 & 6.33 & - \\
HETDEX J141926.25+525441.9 & 14:19:26.25 & +52:54:41.95 & 2.291 & 13.89 & - \\
HETDEX J142017.52+522050.5 & 14:20:17.52 & +52:20:50.55 & 2.298 & 30.32 & - \\
HETDEX J141725.63+523557.5 & 14:17:25.63 & +52:35:57.51 & 2.298 & 9.79 & - \\
HETDEX J141810.58+522031.2 & 14:18:10.58 & +52:20:31.17 & 2.103 & 13.52 & - \\
HETDEX J141826.73+522329.7 & 14:18:26.73 & +52:23:29.71 & 2.308 & 12.46 & - \\
HETDEX J141831.80+522154.0 & 14:18:31.80 & +52:21:53.96 & 2.348 & 12.53 & - \\
HETDEX J141733.68+522437.8 & 14:17:33.68 & +52:24:37.79 & 2.147 & 10.49 & LAE1 \\
HETDEX J141801.61+523101.0 & 14:18:01.61 & +52:31:00.99 & 2.103 & 26.78 & - \\
HETDEX J141802.49+523100.0 & 14:18:02.49 & +52:31:00.01 & 2.103 & 6.96 & - \\
HETDEX J141831.12+523239.7 & 14:18:31.12 & +52:32:39.69 & 2.141 & 22.68 & LAE2 \\
HETDEX J141847.24+523329.5 & 14:18:47.24 & +52:33:29.49 & 2.302 & 9.29 & - \\
HETDEX J142145.41+522401.2 & 14:21:45.41 & +52:24:01.16 & 2.175 & 39.25 & - \\
HETDEX J141852.67+530350.6 & 14:18:52.67 & +53:03:50.64 & 2.254 & 11.19 & - \\
HETDEX J142308.86+525232.6 & 14:23:08.86 & +52:52:32.56 & 2.246 & 20.83 & - \\
HETDEX J141825.85+524355.4 & 14:18:25.85 & +52:43:55.41 & 2.297 & 6.50 & - \\
HETDEX J141834.58+524346.0 & 14:18:34.58 & +52:43:45.97 & 2.188 & 10.69 & LAE3 \\
HETDEX J142144.85+525330.0 & 14:21:44.85 & +52:53:30.01 & 2.341 & 13.29 & - \\
HETDEX J142200.64+525448.7 & 14:22:00.64 & +52:54:48.73 & 2.355 & 9.08 & - \\
HETDEX J141830.26+524329.8 & 14:18:30.26 & +52:43:29.82 & 2.299 & 17.33 & - \\
HETDEX J142026.24+525919.4 & 14:20:26.24 & +52:59:19.36 & 2.289 & 7.90 & - \\
HETDEX J142037.65+530335.6 & 14:20:37.65 & +53:03:35.62 & 2.055 & 17.85 & - 
\enddata 
\centering
\centering
\tablenotetext{^{a}}{
LAEs 1--4 reside in the QSO overdensity, EGS-QO1 (Section \ref{sec: results egs}).
}
\end{deluxetable*}

\subsection{QSO Catalogs} \label{sec: quasar_catalog}
\subsubsection{Foreground QSOs} \label{sec: foreground_qusasar}
The foreground QSOs in our samples are taken from the DR14 QSO catalog \citep[hereafter DR14Q:][]{Paris2018a} of SDSS-IV/eBOSS spectra 
that have a spectral resolution and coverage of $R \approx 2000$ and $3600$--$10400${\AA}, respectively.
In this study, we use foreground QSOs in the cosmic volumes of 
our \hi\ tomography maps (Section \ref{sec: tom map}).
We select foreground QSOs from DR14Q in the redshift range $z=2.05-2.55$ 
in the $0.157$ and $6.0$ deg$^2$ sky areas of the COSMOS and EGS fields, 
and find a total of $0$ and $78$ QSOs, respectively.
Figure \ref{fig: sky egs} presents
the sky distribution of the foreground QSOs in the EGS field.
The basic properties of the foreground QSOs in the EGS field are summarized in Tables \ref{tab: fq egs} and \ref{tab: fq egs cont}.

\begin{table}[t]
\begin{center}
\caption{
Foreground QSOs in the EGS field \label{tab: fq egs} }
\scalebox{0.8}{
\begin{tabular}{cccccc}
\hline \hline 
{ID} &  {R.A.} &  {Decl.} &  {$z_{\rm spec}$} & \\ 
{ }   & {(J2000)} & {(J2000)} & {} &  \\
\hline
7339-56722-0728 & 14:14:16.34 &  +53:35:08.39 & 2.453 \\
7339-56799-0734 & 14:14:20.55 &  +53:22:16.67 & 2.217 \\
7030-56448-0602 & 14:14:22.82 &  +52:51:20.63 & 2.149 \\
7339-56722-0787 & 14:15:24.43 &  +53:28:32.77 & 2.153 \\
7339-56799-0238 & 14:15:34.20 &  +52:57:43.22 & 2.061 \\
7340-56837-0794 & 14:15:41.15 &  +53:51:04.20 & 2.420 \\
7339-56768-0256 & 14:15:48.07 &  +52:09:09.94 & 2.469 \\
7339-56799-0787 & 14:15:54.32 &  +53:53:57.02 & 2.191 \\
7339-56722-0788 & 14:15:54.46 &  +53:17:06.92 & 2.138 \\
7339-56799-0770 & 14:16:02.71 &  +53:17:45.03 & 2.207 \\
6717-56397-0604 & 14:16:27.00 &  +53:19:40.10 & 2.428 \\
7339-56799-0809 & 14:16:28.69 &  +53:31:00.40 & 2.273 \\
7029-56455-0247 & 14:16:28.92 &  +52:03:29.00 & 2.134 \\
7339-56722-0838 & 14:16:41.41 &  +53:21:47.17 & 2.214 \\
7028-56449-0809 & 14:16:45.06 &  +53:05:10.15 & 2.529 \\
7339-56772-0218 & 14:16:47.20 &  +52:11:15.26 & 2.158 \\
7338-56745-0823 & 14:17:04.00 &  +53:38:07.47 & 2.501 \\
7339-56799-0194 & 14:17:15.19 &  +53:03:03.76 & 2.164 \\
7028-56449-0805 & 14:17:22.72 &  +52:58:51.62 & 2.405 \\
7339-56751-0060 & 14:17:26.51 &  +52:18:56.51 & 2.151 \\
7028-56449-0834 & 14:17:29.99 &  +53:38:25.69 & 2.119 \\
7339-56722-0200 & 14:17:38.83 &  +52:23:33.07 & 2.153 \\
7339-56722-0832 & 14:17:43.33 &  +53:11:45.67 & 2.059 \\
7339-56799-0831 & 14:17:50.37 &  +53:45:17.76 & 2.177 \\
7339-56772-0798 & 14:17:52.39 &  +53:48:49.43 & 2.093 \\
7339-56799-0854 & 14:18:07.73 &  +53:17:54.02 & 2.278 \\
7339-56722-0876 & 14:18:17.46 &  +53:11:16.82 & 2.232 \\
7339-57518-0151 & 14:18:18.45 &  +52:43:56.05 & 2.136 \\
7029-56455-0234 & 14:18:23.07 &  +52:41:18.81 & 2.050 \\
7338-56745-0149 & 14:18:42.27 &  +52:36:43.97 & 2.128 \\
7339-56772-0893 & 14:18:43.30 &  +53:19:20.83 & 2.301 \\
7030-56448-0306 & 14:18:57.23 &  +52:18:23.39 & 2.167 \\
7339-56772-0895 & 14:19:05.24 &  +53:53:54.17 & 2.427 \\
7339-56799-0134 & 14:19:05.73 &  +52:12:38.07 & 2.219 \\
7031-56449-0404 & 14:19:07.20 &  +52:01:51.74 & 2.172 \\
7340-56825-0873 & 14:19:10.22 &  +53:47:07.11 & 2.373 \\
7028-56449-0870 & 14:19:15.99 &  +53:49:24.13 & 2.209 \\
7339-56772-0889 & 14:19:27.35 &  +53:37:27.70 & 2.368 \\
7339-56772-0884 & 14:19:29.90 &  +53:35:01.41 & 2.390 \\
7339-56799-0105 & 14:19:32.07 &  +52:26:39.46 & 2.162 \\
7028-56449-0101 & 14:19:45.40 &  +52:23:33.57 & 2.378 \\
7339-56799-0087 & 14:19:52.89 &  +52:01:16.87 & 2.229 \\
7339-56799-0106 & 14:19:55.27 &  +52:27:41.19 & 2.141 \\
7339-56722-0093 & 14:20:36.56 &  +52:14:55.05 & 2.212 \\
7028-56449-0937 & 14:20:37.24 &  +52:58:51.00 & 2.274 \\
7340-56837-0923 & 14:20:41.26 &  +53:33:55.30 & 2.421 \\
7028-56449-0066 & 14:20:46.11 &  +52:24:21.61 & 2.256 \\
7339-56799-0913 & 14:20:49.31 &  +53:52:11.59 & 2.221 \\
7029-56455-0158 & 14:20:58.63 &  +52:40:44.43 & 2.489 \\
7339-56772-0955 & 14:21:02.17 &  +53:39:44.14 & 2.292 \\
\hline
\end{tabular}}
\end{center}
\end{table}

\begin{table}[t]
\begin{center}
\caption{
Foreground QSOs in the EGS field (continued) \label{tab: fq egs cont} }
\scalebox{0.8}{
\begin{tabular}{cccccc}
\hline
{ID} &  {R.A.} &  {Decl.} &  {$z_{\rm spec}$} \\ 
{ }   & {(J2000)} & {(J2000)} & {} \\
\hline \hline 
7028-56449-0067 & 14:21:03.96 &  +52:37:12.53 & 2.235 \\
7339-56722-0074 & 14:21:17.99 &  +52:53:46.00 & 2.308 \\
7028-56449-0933 & 14:21:33.92 &  +53:02:45.52 & 2.150 \\
7339-56722-0062 & 14:21:55.20 &  +52:27:49.48 & 2.516 \\
7339-56799-0038 & 14:22:01.46 &  +52:32:50.26 & 2.121 \\
7339-56799-0037 & 14:22:08.12 &  +52:29:08.65 & 2.370 \\
7029-56455-0898 & 14:22:26.24 &  +52:57:09.93 & 2.095 \\
7340-56726-0034 & 14:22:34.46 &  +52:58:38.02 & 2.138 \\
7030-56448-0218 & 14:22:34.99 &  +52:00:10.05 & 2.109 \\
7339-56772-0038 & 14:22:37.49 &  +52:53:35.86 & 2.226 \\
7032-56471-0332 & 14:22:40.47 &  +52:04:11.81 & 2.267 \\
7339-56780-0074 & 14:22:42.59 &  +52:44:15.69 & 2.171 \\
7339-56772-0868 & 14:22:52.42 &  +53:36:48.86 & 2.084 \\
7340-56837-0978 & 14:23:06.05 &  +53:15:29.03 & 2.468 \\
7028-56449-0945 & 14:23:07.38 &  +53:34:39.84 & 2.074 \\
7029-56455-0086 & 14:23:33.95 &  +52:07:00.95 & 2.271 \\
7339-57481-0991 & 14:23:37.51 &  +53:18:28.89 & 2.435 \\
7339-56799-0984 & 14:23:50.24 &  +53:29:29.31 & 2.136 \\
7339-56799-0992 & 14:24:11.08 &  +53:20:41.38 & 2.362 \\
7339-56772-0972 & 14:24:19.18 &  +53:17:50.62 & 2.530 \\
7339-56768-0016 & 14:24:22.50 &  +52:59:03.22 & 2.138 \\
7032-56471-0306 & 14:24:27.85 &  +52:20:44.40 & 2.331 \\
7029-56455-0032 & 14:24:32.08 &  +52:22:20.49 & 2.194 \\
7031-56449-0346 & 14:24:38.98 &  +52:21:39.15 & 2.259 \\
7031-56449-0655 & 14:24:48.10 &  +53:21:21.42 & 2.066 \\
7030-56448-0159 & 14:25:06.97 &  +52:54:44.33 & 2.546 \\
7032-56471-0723 & 14:25:23.43 &  +53:29:45.88 & 2.182 \\
7032-56471-0298 & 14:25:51.03 &  +52:05:09.06 & 2.315 \\
\hline
\end{tabular}}
\end{center}
\end{table}

\subsubsection{Background QSOs} \label{sec: background_qusasar}
The background QSOs are also taken from the DR14Q catalog.
We only make a sample of background QSOs in the EGS field.
This is because we do not need to use background QSOs for 
the \hi\ tomography map in the COSMOS field, where a high-resolution \hi\ tomography map is already available (Section \ref{subsec: tom cosmos}).
We select background QSOs from DR14Q in the redshift range $z=2.1$--$3.1$ 
in the $6.0$ \sqdeg\ sky area of the EGS field. 
The redshift range of $z=2.1$--$3.1$ is chosen, 
because we aim to investigate the \hi\ Ly$\alpha$ forest of foreground absorbers in the same redshift range as those of the COSMOS \hi\ tomography map 
($z = 2.05$--$2.55$; Section \ref{subsec: tom cosmos}).
We apply these two criteria, and obtain $128$ background QSOs. 
\par
For our analysis of \hi\ Ly$\alpha$ forest absorption, 
we investigate these QSO spectra, and conduct further selection.
We apply a criterion that QSO spectra should have a median signal-to-noise ratio (S/N) $\geq 2$ per pixel over their \lya\ forest wavelength range (i.e., $1041$--$1185${\AA} in the rest-frame; \citealt{Mukae2017a}). 
In addition, we remove QSOs whose spectra have broad absorption lines 
whose BALnicity Index (BI) blueward of \Civ\ emission is BI $<200$ km s$^{-1}$ 
in the DR14Q catalog. We also remove QSOs with a damped \lya\ system (DLA) 
in the \lya\ forest wavelength range on the basis of the DLA catalog of \cite{Noterdaeme2012a} and their updated one\footnote{ \url{http://www2.iap.fr/users/noterdae/DLA/DLA.html} }
for the SDSS DR12 QSOs \citep{Paris2017a}.
For QSOs that have no SDSS DR12 counterpart, 
we visually inspect the QSO spectra, and 
examine whether signatures of DLAs exist in the \lya\ forest wavelength range.
\par 
Our selection gives a total of $43$ background QSOs for the \hi\ tomography analysis. 
The distribution of the background QSOs is shown in Figure \ref{fig: sky egs}. The basic properties of the $43$ background QSOs are summarized in Table \ref{tab: bq egs}. Some example QSO spectra are shown in Figure \ref{fig: bq spec}.

\begin{table}[t]
\begin{center}
\caption{
Background QSOs in the EGS field  \label{tab: bq egs} }
\scalebox{0.8}{
\begin{tabular}{cccccc}
\hline \hline 
{ID} &  {R.A.} &  {Decl.} &  {$z_{\rm spec}$} & {$g$}\\ 
{ }   & {(J2000)} & {(J2000)} & {} & {(AB)} \\
\hline
7339-56799-0270 & 14:14:08.64 & +52:40:38.64 & 2.790 & 20.59 \\
7339-56722-0800 & 14:14:18.24 & +53:50:46.68 & 2.729 & 21.43 \\
7339-56799-0734 & 14:14:20.64 & +53:22:16.68 & 2.212 & 19.54 \\
7339-56799-0730 & 14:14:35.52 & +53:25:36.84 & 2.861 & 20.72 \\
7339-56722-0297 & 14:14:39.12 & +52:06:16.20 & 2.914 & 20.93 \\
7339-56799-0728 & 14:14:44.16 & +53:35:55.68 & 2.734 & 21.55 \\
7339-56799-0277 & 14:15:08.64 & +53:00:19.80 & 2.765 & 21.36 \\
7340-56837-0794 & 14:15:41.04 & +53:51:04.32 & 2.420 & 20.81 \\
7027-56448-0068 & 14:15:51.36 & +52:27:40.68 & 2.583 & 19.86 \\
7028-56449-0809 & 14:16:45.12 & +53:05:10.32 & 2.529 & 21.68 \\
7339-56772-0218 & 14:16:47.28 & +52:11:15.36 & 2.153 & 18.70 \\
7340-56837-0833 & 14:17:22.32 & +53:48:52.92 & 2.726 & 20.71 \\
7339-56799-0831 & 14:17:50.40 & +53:45:17.64 & 2.190 & 20.30 \\
7339-56799-0854 & 14:18:07.68 & +53:17:53.88 & 2.274 & 20.67 \\
7339-56722-0876 & 14:18:17.52 & +53:11:16.80 & 2.238 & 20.50 \\
7339-56772-0893 & 14:18:43.20 & +53:19:21.00 & 2.298 & 20.56 \\
7340-56837-0117 & 14:19:12.48 & +52:08:17.88 & 2.563 & 19.78 \\
7339-56799-0087 & 14:19:52.80 & +52:01:17.04 & 2.224 & 19.48 \\
7339-56772-0924 & 14:20:10.56 & +53:12:23.76 & 2.597 & 20.09 \\
7027-56448-0994 & 14:20:33.12 & +53:07:35.04 & 2.880 & 21.55 \\
7340-56837-0923 & 14:20:41.28 & +53:33:55.44 & 2.421 & 20.87 \\
7339-56799-0074 & 14:21:13.20 & +52:49:30.00 & 2.644 & 19.36 \\
7339-56722-0074 & 14:21:18.00 & +52:53:45.96 & 2.306 & 19.95 \\
7339-56799-0069 & 14:21:38.64 & +52:33:24.48 & 2.606 & 20.30 \\
7339-56799-0068 & 14:21:41.28 & +52:45:51.84 & 2.654 & 21.04 \\
7339-56722-0062 & 14:21:55.20 & +52:27:49.32 & 2.516 & 21.10 \\
7339-56768-0038 & 14:22:39.60 & +52:28:52.68 & 2.989 & 21.61 \\
7339-56780-0074 & 14:22:42.48 & +52:44:15.72 & 2.175 & 20.03 \\
7340-56837-0978 & 14:23:06.00 & +53:15:29.16 & 2.468 & 18.29 \\
7029-56455-0100 & 14:23:17.28 & +52:13:12.72 & 2.671 & 21.54 \\
7032-56471-0340 & 14:23:37.20 & +52:16:07.68 & 2.894 & 19.73 \\
7339-56799-0992 & 14:24:11.04 & +53:20:41.28 & 2.366 & 20.31 \\
6710-56416-0442 & 14:24:11.52 & +53:50:26.88 & 2.769 & 20.96 \\
7339-56799-0014 & 14:24:18.24 & +53:04:06.60 & 2.859 & 19.56 \\
7339-56772-0972 & 14:24:19.20 & +53:17:50.64 & 2.530 & 20.49 \\
7032-56471-0306 & 14:24:27.84 & +52:20:44.52 & 2.324 & 20.03 \\
7029-56455-0955 & 14:24:33.12 & +53:43:52.68 & 2.711 & 20.51 \\
7030-56448-0160 & 14:24:50.88 & +52:50:01.68 & 2.728 & 19.71 \\
7030-56448-0130 & 14:24:55.68 & +52:06:09.72 & 2.631 & 21.20 \\
7031-56449-0356 & 14:24:58.56 & +52:41:49.92 & 3.015 & 21.27 \\
7031-56449-0334 & 14:25:06.24 & +52:01:28.92 & 2.736 & 21.92 \\
6710-56416-0446 & 14:25:09.12 & +53:51:49.32 & 3.102 & 21.37 \\
7032-56471-0729 & 14:25:10.80 & +53:23:09.60 & 2.861 & 20.73 \\
\hline
\end{tabular}}
\end{center}
\end{table}

\section{\hi\ TOMOGRAPHY TECHNIQUES AND MAPS}\label{sec: tom tech and map}
We carry out \hi\ tomography mapping that is a technique 
to reconstruct the 3D \hi\ LSSs based on \hi\ absorption features 
found in the sightlines to multiple background source spectra \citep[e.g.,][]{Lee2018a, Lee2014b, Lee2014a, Caucci2008a, Pichon2001a}. 
This Section describes 
how we make \hi\ tomography maps from background source spectra (Section \ref{sec: tom tech}),
and presents \hi\ tomography maps of the COSMOS and EGS fields 
(Section \ref{sec: tom map}).
Note that we make a new \hi\ tomography map only in the EGS field.
This is because we use the public data of the \hi\ tomography map 
in the COSMOS field \citep[][Section \ref{subsec: tom cosmos}]{Lee2018a}.

\subsection{\hi\ Tomography Techniques} \label{sec: tom tech}
Our \hi\ tomography analysis consists of the following two processes:
1) normalizing the background source spectra 
with estimated continua to create input spectra for \hi\ tomography (Section \ref{subsec: intrinsic continua})
and 
2) reconstructing the \hi\ LSSs from the normalized spectra 
(Section \ref{subsec: tomographic reconstruction}). 

\subsubsection{Intrinsic Continua} \label{subsec: intrinsic continua}
To probe \hi\ absorption along the lines of sight to the background QSOs,
we estimate the \lya\ forest transmission in the $1041$--$1185${\AA} rest frame,
\begin{equation}
F (z) = f_{\rm obs} / f_{\rm int}, 
\label{eq:Fz}
\end{equation}
where $f_{\rm obs}$ is the observed continuum flux density 
and  $f_{\rm int}$ is the intrinsic continuum flux density 
that is not affected by the Ly$\alpha$ forest absorption due to the IGM.
We estimate $f_{\rm int}$ of our background QSOs, 
applying the continuum fitting technique of 
the mean-flux regulated/principal component analysis 
(MF-PCA; \citealt{Lee2012a})
with the code developed by \citeauthor{Lee2013a} 
(\citeyear{Lee2013a}; see also \citealt{Lee2014b}).
In this technique, there are two steps.
The first step is to fit spectral templates of QSOs 
to the observed spectra redward of Ly$\alpha$ 
to obtain initial estimates of 
the continuum spectra blueward of Ly$\alpha$. 
In the same manner as \cite{Lee2014b},
we use the spectral templates of QSOs constructed by \cite{Suzuki2005a}. 
The second step is to constrain 
the amplitude and slope of the blueward spectra 
that should match to previous measurements of 
the cosmic mean \lya\ forest transmission, ${F_{\rm cos} (z)}$.  
We adopt ${F_{\rm cos} (z)}$ estimated by \cite{Faucher-Giguere2008a}, 
\begin{eqnarray}
	{F_{\rm cos} (z)}=\exp[-0.00185(1+z)^{3.92}].
	\label{eq:mean flux}
\end{eqnarray}
\par
\begin{figure}[ht]
 \centering
 \includegraphics[width=1.1\hsize, clip, bb= 0 0 800 800, clip=true]{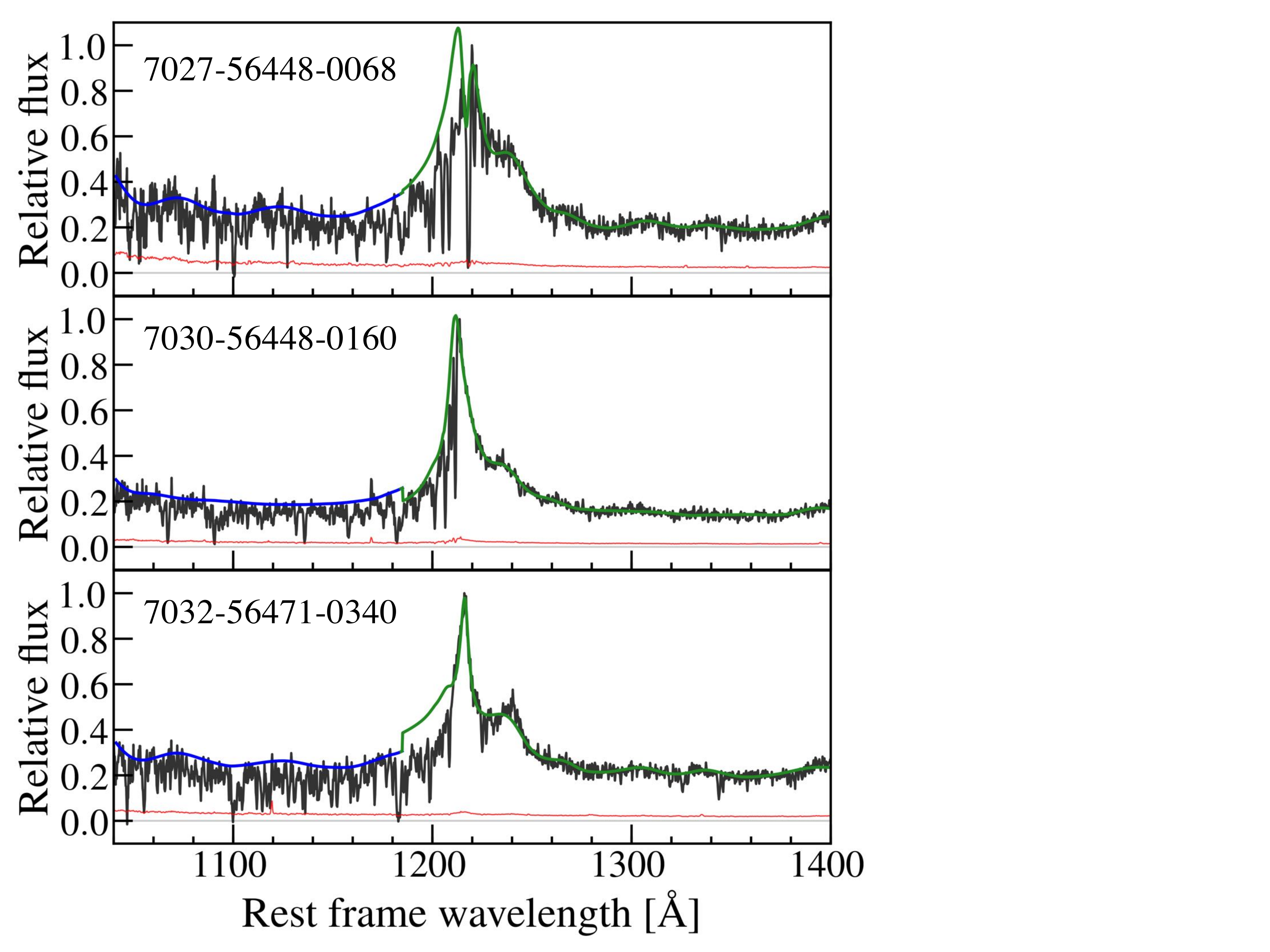}
 \caption{
 Example spectra of our background QSOs. 
 The black (red) lines depict flux (error) per pixel.
 The grey lines show zero flux level.
 The green curves are the \cite{Suzuki2005a}
 template overplotted at the source redshifts.
 The blue curves are the \cite{Suzuki2005a}
 template representing the estimated continua.
 }
 \label{fig: bq spec}
\end{figure}

With $F_{\rm cos}$ we estimate $f_{\rm int}$ (Figure \ref{fig: bq spec}), and use Equation (\ref{eq:Fz}) to obtain $F(z)$.  
Note that the strong stellar and interstellar absorptions 
of N{\sc ii} $\lambda 1084$ and C{\sc iii} $\lambda 1175$ 
associated with the QSO host galaxies 
in the \lya\ forest wavelength range could bias the results. 
For conservative estimates, we do not use the spectra 
in the wavelength ranges of $\pm5${\AA} around these lines in our analyses.
The uncertainties of $F(z)$ are calculated  
from the errors of the $f_{\rm obs}$ measurements 
and the $f_{\rm int}$ estimates based on the MF-PCA continuum fitting, 
the latter of which are evaluated by \cite{Lee2012a} 
as a function of redshift and median S/N over 
the \lya\ forest wavelength range (see Figure 8 of \citealt{Lee2012a}).  
Specifically, we adopt MF-PCA continuum fitting errors of 
$7${\%}, $6${\%}, and $4${\%} for spectra with median S/Ns 
over the \lya\ forest wavelength ranges of $2$--$4$, $4$--$10$, and $>10$, respectively.  
\par 
Based on the estimated $F(z)$ and the cosmic mean Ly$\alpha$ forest transmission $F_{\rm cos}(z)$, we calculate 
the Ly$\alpha$ forest fluctuation 
(hereafter referred to as \hi\ transmission overdensity)
$\delta_{F}$ for our background QSOs, 
\begin{eqnarray}
    \delta_{F} = \frac{F(z)}{F_{\rm cos} (z)}-1,
    \label{eq: transmission overdensity}
\end{eqnarray}
where negative values correspond to strong \hi\ absorptions. 
The errors of $\delta_F$ are calculated 
with the uncertainties of $F(z)$. 
We confirm that the systematic effect of 
using different prescriptions of $F_{\rm cos}(z)$ 
obtained by \cite{Becker2013a} and \cite{Inoue2014a} is minor, 
only within $2${\%}, 
which is not as large as the uncertainties of $F(z)$. 

\subsubsection{Reconstruction Processes} \label{subsec: tomographic reconstruction}
Once we obtain the $\delta_{F}$ spectra of the background QSOs, 
we carry out \hi\ tomographic reconstruction 
to reveal the 3D distribution of the \hi\ gas. 
In the same manner as \cite{Lee2018a, Lee2016a, Lee2014b},
we use the reconstruction code developed by \cite{Stark2015a}.\footnote{ \url{https://github.com/caseywstark/dachshund} } 
The reconstruction code performs Wiener filtering for the estimated $\delta_F$ values along the sightlines of our background QSOs.
The Wiener filtering is based on the following two calculations:
the first is Gaussian smoothing with the scale of 
the mean transverse sightline separation \dperp, 
which determines the spatial resolution of the tomography map; 
the second is input pixel weighting by the S/N to remove possible systematics caused by low S/N spectra.
\par 
Specifically, in the Wiener filtering, 
the reconstructed \hi\ transmission overdensity map $\delrecon$ 
is given by the following estimator
\citep{Stark2015a, Lee2014a, Caucci2008a, Pichon2001a},
\beq \label{eq:wiener} \delrecon=\cmd\cdot (\cdd+\mathbf{N})^{-1}\cdot \delobson, \eeq
where \delobson\ is the input  \hi\ transmission overdensity datacube comprised of our background source spectra and coordinates.
$\cmd$, $\cdd$, and $\mathbf{N}$ are the map-datacube, datacube-datacube, and noise covariances, respectively. 
The estimator, $\cmd\cdot (\cdd+\mathbf{N})^{-1}$, is constructed 
so that it minimizes an expected error between the reconstructed $\delrecon$ and the actual \hi\ distribution \cite[][and references therein]{Stark2015a}.
The estimator also allows us to down-weight pixels in low S/N spectra.
These covariances are assumed to be a Gaussian covariance 
between any two points $\mathbf{r_1}$ and $\mathbf{r_2}$.
\beq \cdd=\cmd=\mathbf{C(r_1,r_2)} \eeq and 
\beq\mathbf{C(r_1,r_2)}=\sigma_F^2\exp\left[-\frac{(\Delta r_\parallel)^2}{2L^2_\parallel}\right] \exp\left[-\frac{(\Delta r_\perp)^2}{2L^2_\perp}\right],\eeq
where $\Delta r_\parallel$ and $\Delta r_\perp$ ($L_\parallel$ and $L_\perp$) are the line-of-sight (LOS) and transverse distances
between $\mathbf{r_1}$ and $\mathbf{r_2}$ (correlation lengths), respectively. 
We adopt $L_\parallel=L_\perp=\langle d_{\perp }\rangle$ as well as a normalization of $\sigma_F^2= 0.05$ 
in the same manner as \cite{Lee2018a, Lee2014b}.
More details about the reconstruction process is presented in \cite{Stark2015a} and \cite{Lee2018a}.

\subsection{\hi\ Tomography Maps} \label{sec: tom map}
In the COSMOS field, we use the \hi\ tomography map of \cite{Lee2018a} (Section \ref{subsec: tom cosmos}), while in the EGS field 
we make the \hi\ tomography map with the $\delta_{F}$ spectra 
of our background QSOs (Section \ref{subsec: tom egs}). 

\subsubsection{COSMOS} \label{subsec: tom cosmos}
\begin{figure*}[p]
\begin{tabular}{c}
\begin{minipage}[t]{1,0\hsize}
\centering
\includegraphics[width=0.8\hsize, clip, bb=0 0 900 600, clip=true]{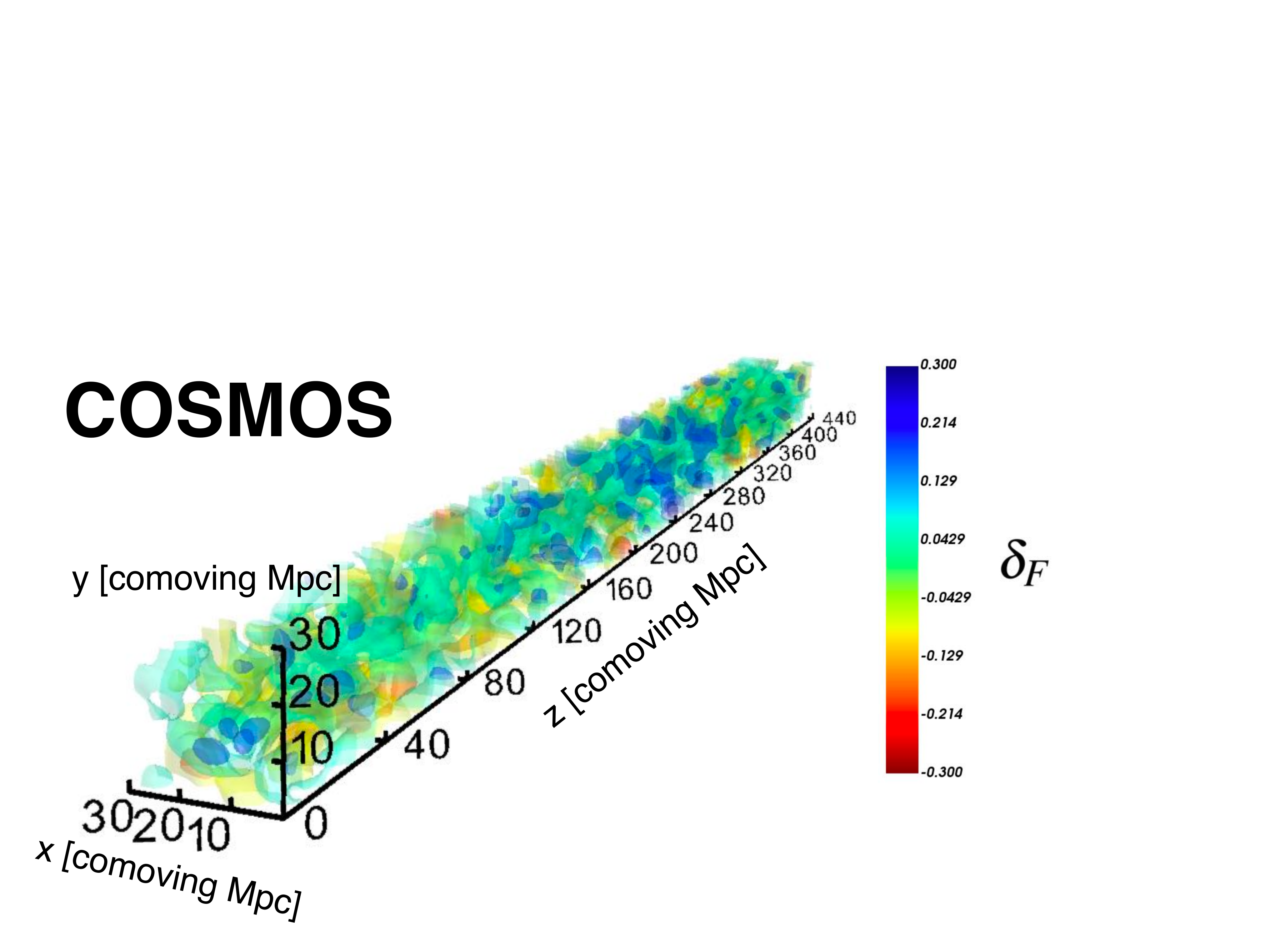}
 \caption{
COSMOS \hi\ tomography map reconstructed from \hi\ absorption in the spectra of the background galaxies and QSOs \citep{Lee2018a}. The spatial axes of R.A., Decl., and $z$ (Hubble flow distances)
correspond to the $x$, $y$, and $z$ axes, respectively, that are shown in comoving scale. The redshift range of the \hi\ tomography map is $z=2.05$--$2.55$. The color contours represent the \hi\ transmission overdensity $\delta_{F}$ whose negative values (in red color) correspond to high \hi\ overdensities. The $\delta_{F}$'s maximum (minimum) scale of this figure is set to $+0.3$ ($-0.3$) for visualization. The $\delta_F$ values of some volumes do not fall in the range of $-0.3 < \delta_F < 0.3$, but 
all are 
in $-0.5 < \delta_F < 0.5$.
The cosmic large-scale structures are roughly traced by the \hi\ transmission overdensities.
}
 \label{fig: hi_map_cosmos}
\end{minipage}
\\
\begin{minipage}[t]{1,0\hsize}
\centering
\includegraphics[width=1.0\hsize, clip, bb= 0 0 1100 500, clip=true]{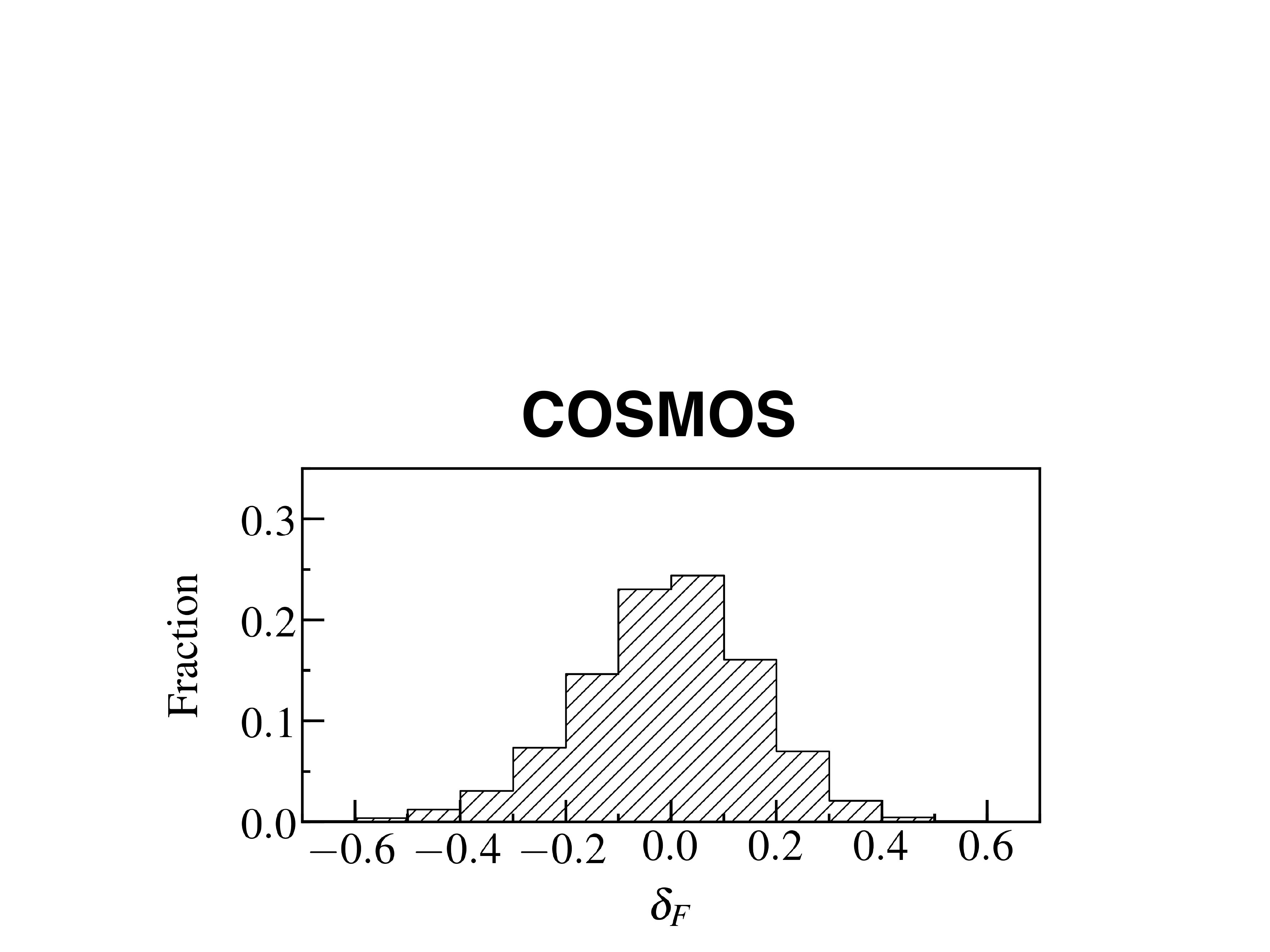}
\caption{
 Histogram of the \dF\ pixel distribution of the COSMOS \hi\ tomography map.
 }
\label{fig: hist_delF_cosmos}
\end{minipage}
\end{tabular}
\end{figure*}

For the COSMOS field,
we use the public data of the COSMOS \hi\ tomography map 
made by the CLAMATO survey \citep{Lee2018a}
\footnote{ \url{https://clamato.lbl.gov/} }.
The COSMOS \hi\ tomography map is a 3D map of the IGM \hi\ absorption 
at $z=2.05$--$2.55$ in a $0.157$ \sqdeg\ area of the COSMOS field, 
having a $30 \times\ 24 \times\ 444\ h^{-3} {\rm cMpc}^3$ cosmic volume 
with a spatial resolution of $2.5$ \hmpc\ and a grid size of $0.5$ \hmpc.
The \hi\ tomography map is reconstructed from the spectra of
240 background galaxies and QSOs at $z= 2.2$--$3.0$.
The sky distribution of these background sources is shown in Figure \ref{fig: sky cosmos}.
Figures \ref{fig: hi_map_cosmos} and \ref{fig: hist_delF_cosmos} present the COSMOS \hi\ tomography map and the \dF\ pixel distribution, respectively.
For error estimates in Section \ref{sec: results cosmos}, 
we generate 1000 mock \hi\ tomography maps 
to which we give random perturbations following the Gaussian distribution 
with sigma defined by the data values of the error map, 
where the error map is estimated from the $1\sigma$ uncertainties of $\delta_F$ spectra of the background sources in the CLAMATO survey \citep{Lee2018a}.
The typical 1$\sigma$ uncertainty of $\delta_F$ for a pixel in the \hi\ tomographic map is found to be about 0.1.

\subsubsection{EGS} \label{subsec: tom egs}
\begin{figure*}[p]
\begin{tabular}{c}
\begin{minipage}[t]{1,0\hsize}
\centering
\includegraphics[width=0.8\hsize,  clip, bb=0 0 900 600, clip=true]{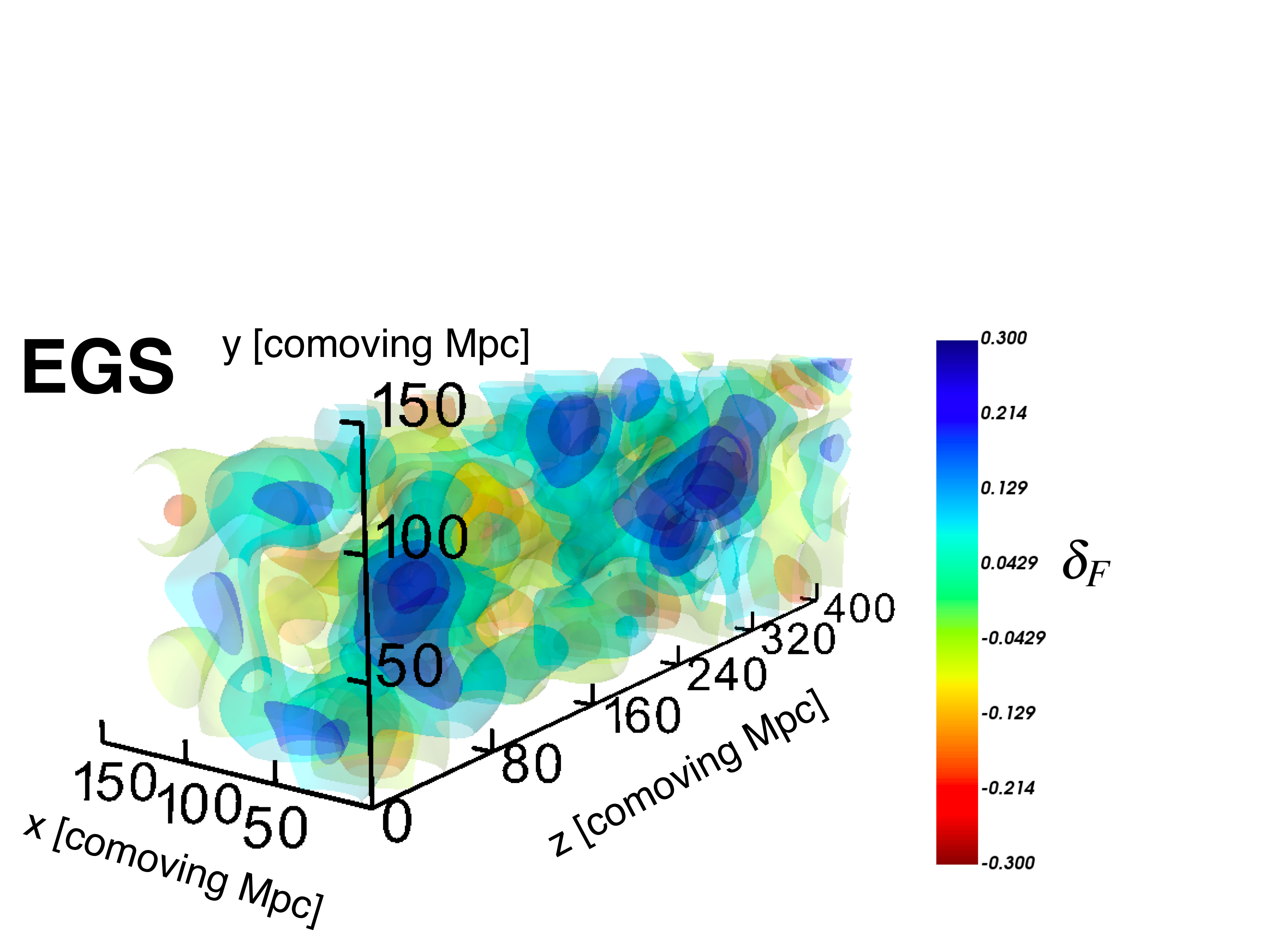}
 \caption{
 Same as Figure \ref{fig: hi_map_cosmos}, but for the EGS \hi\ tomography map. Note that the spatial resolution is $20$ \hmpc\, which is larger than the one of COSMOS ($2.5$ \hmpc). The $\delta_F$ values of some volumes do not fall in the range of $-0.3 < \delta_F < 0.3$, but all are
 in $-0.6 < \delta_F < 0.6$. 
 This \hi\ tomography map does not have a spatial resolution 
  as high as the one of Figure \ref{fig: hi_map_cosmos}, 
  but covers a cosmic volume larger than 
  that of Figure \ref{fig: hi_map_cosmos}.}
 \label{fig: hi_map_egs}
\end{minipage}
\\
\begin{minipage}[t]{1,0\hsize}
\centering
\includegraphics[width=1.0\hsize, clip, bb= 0 0 1100 500, clip=true]{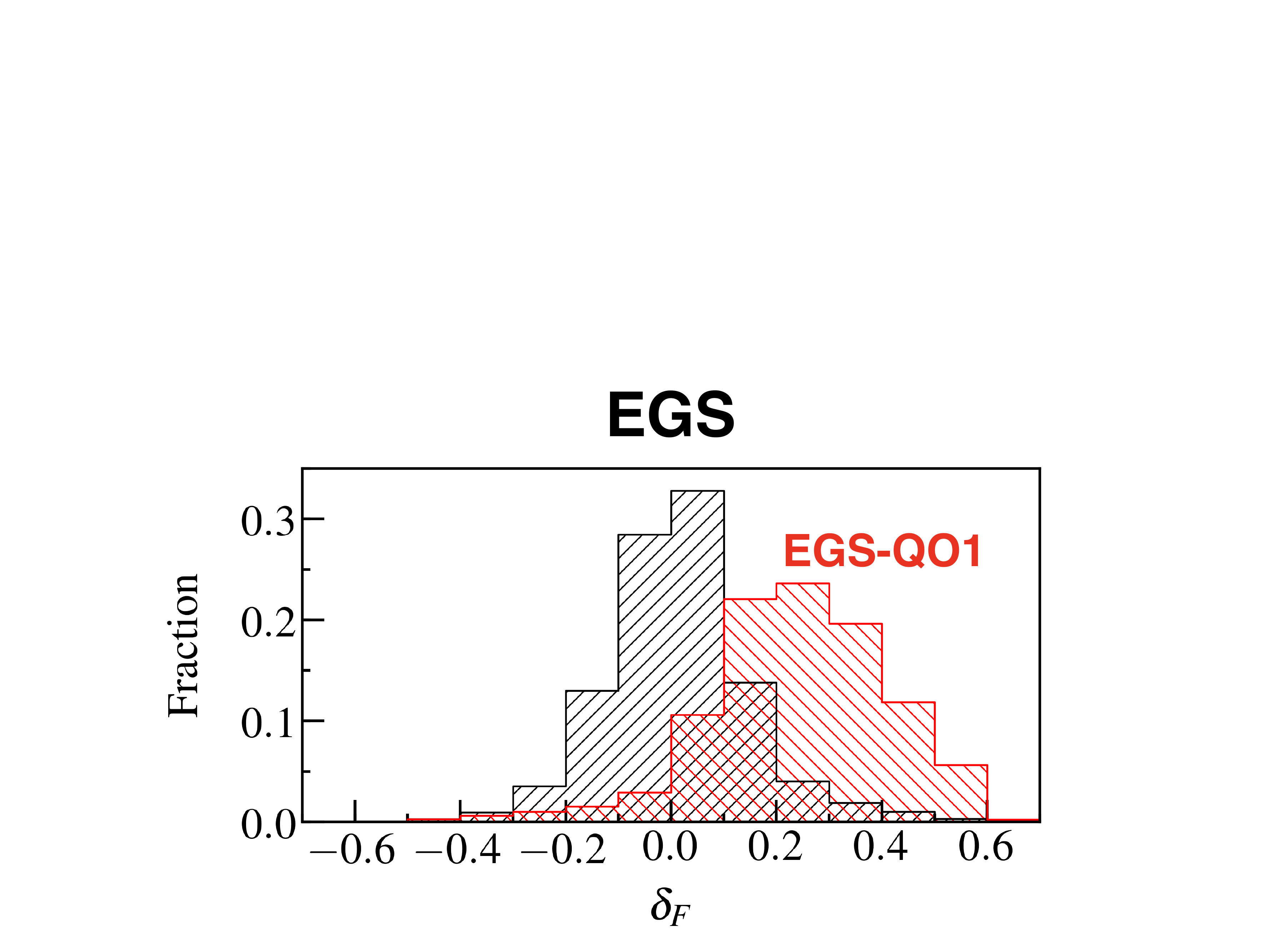}
\caption{
 Same as Figure \ref{fig: hist_delF_cosmos}, but for the EGS \hi\ tomography map.
 The red histogram presents the \dF\ pixel distribution of the QSO overdensity EGS-QO1 (Section \ref{sec: results egs}).
 }
\label{fig: hist_delF_egs}
\end{minipage}
\end{tabular}
\end{figure*}

For the EGS field,  we conduct large-scale \hi\ tomography mapping 
with our background QSOs (Figure \ref{fig: sky egs}).
The mean transverse sightline separation is \dperp\ $=20$ \hmpc\ 
which is about ten times larger than those of the background sources 
in the COSMOS field (Section \ref{subsec: tom cosmos}).
We aim to complement the COSMOS-field \hi\ tomography 
with the EGS-field \hi\ tomography that covers a large volume, 
albeit with coarse resolution.
For our tomographic reconstruction, we choose a redshift range of 
$z = 2.05$--$2.55$ that is the same as the one of the COSMOS \hi\
tomography map \citep[][Section \ref{subsec: tom cosmos}]{Lee2018a}.
This redshift range and the $6.0$ \sqdeg\ sky area of the EGS field 
give an overall cosmic volume of 
$124\ \times 136\ \times 444\ h^{-3} {\rm cMpc}^{3}$. 
We adopt a grid size of $1.0\ h^{-1}$ cMpc that 
over-samples the spatial resolution of $20$ \hmpc.
Figures \ref{fig: hi_map_egs} and \ref{fig: hist_delF_egs} present the EGS \hi\ tomography map and the \dF\ pixel distribution, respectively.
For error estimates, we create mock Ly$\alpha$ forest transmission data 
$F(z)$ for the $43$ background sightlines, 
adding Gaussian-distribution random noise based on the uncertainties of
$F(z)$ that are obtained in Section \ref{subsec: intrinsic continua}. 
We then perform \hi\ tomography mapping with the mock data. Repeating this process to produce 100 mock \hi\ tomography maps, with the total limited by computing resources.
The typical 1$\sigma$ uncertainty of $\delta_F$ for a pixel in the \hi\ tomographic map is found to be about 0.1.
\par 
One might expect that 
the \hi-gas distribution in the EGS \hi\ tomography map
could be affected by interpolation in the tomographic reconstruction process, due to the coarse distribution of sightlines. 
Recently, \cite{Ravoux2020a} have simulated  
large-scale \hi\ tomography of eBOSS background QSOs 
whose sightline separation is $\simeq15$ \hmpc, and 
demonstrated that the correlation of the reconstructed \hi-gas distribution 
and the underlying matter field is retained on large-scales.
The correlation of \hi-gas and underlying matter is also investigated by \cite{Cai2016a} and a strong correlation is suggested on $15-25$ \hmpc\ scales
\footnote{
It is noted that the \hi-gas distribution on a few Mpc scales
can be affected by the strong ionizing radiation of the QSOs \citep{Mukae2020a, Momose2020b}}.
The simulation study of \cite{Ozbek2016a} assessed 
the statistical properties of large-scale \hi\ tomography 
based on eBOSS background QSOs. 
They estimated the root-mean-square error for a $\delta_F$ pixel 
reconstructed with $\simeq20$ \hmpc\ resolution is $\simeq0.02$
which is smaller than the $\delta_F$ pixel error of the EGS \hi\ tomography map.

\section{Results} \label{sec: results}
In this section,  we investigate the IGM \hi-gas distributions around $z\sim2$ galaxies in our two \hi\ tomography maps.

\subsection{Spatial Correlations Between \hi\ Gas and Galaxies} \label{sec: results cosmos}
\par 
We present the results of spatial correlations between IGM \hi\ gas and galaxies in the COSMOS blank field. 
The results consist of two components:
a 2D distribution map of \dF\ as a function of distance and
a radial profile of \dF\ (hereafter \hrp s) as a function of distance.
\par 
We measure \dF\ around the 27 LAEs (Section \ref{sec: lae_catalog}) in our COSMOS \hi\ tomography map (Section \ref{subsec: tom cosmos}) along the transverse $D$ and the line-of-sight (LOS) $Z$ directions. 
The comoving distances of $D$ and $Z$ are computed under the assumption of Hubble flow.
We take the average of \hi\ tomography pixel arrays over a distance $\pm$Z/2 \hmpc\ along the LOS direction at a fixed distance D \hmpc\ in the transverse direction from LAEs. 
Here we use \lya\ redshifts for the redshifts of LAEs. 
We estimate $1\sigma$ errors of the averaged \dF, calculating standard deviations of the measurements with the 1000 mock tomography maps (Section \ref{subsec: tom cosmos}).
\par
\begin{figure*}[h]
\centering
\includegraphics[width=1.0\hsize, clip, bb = 100 0 1000 550, clip=true]{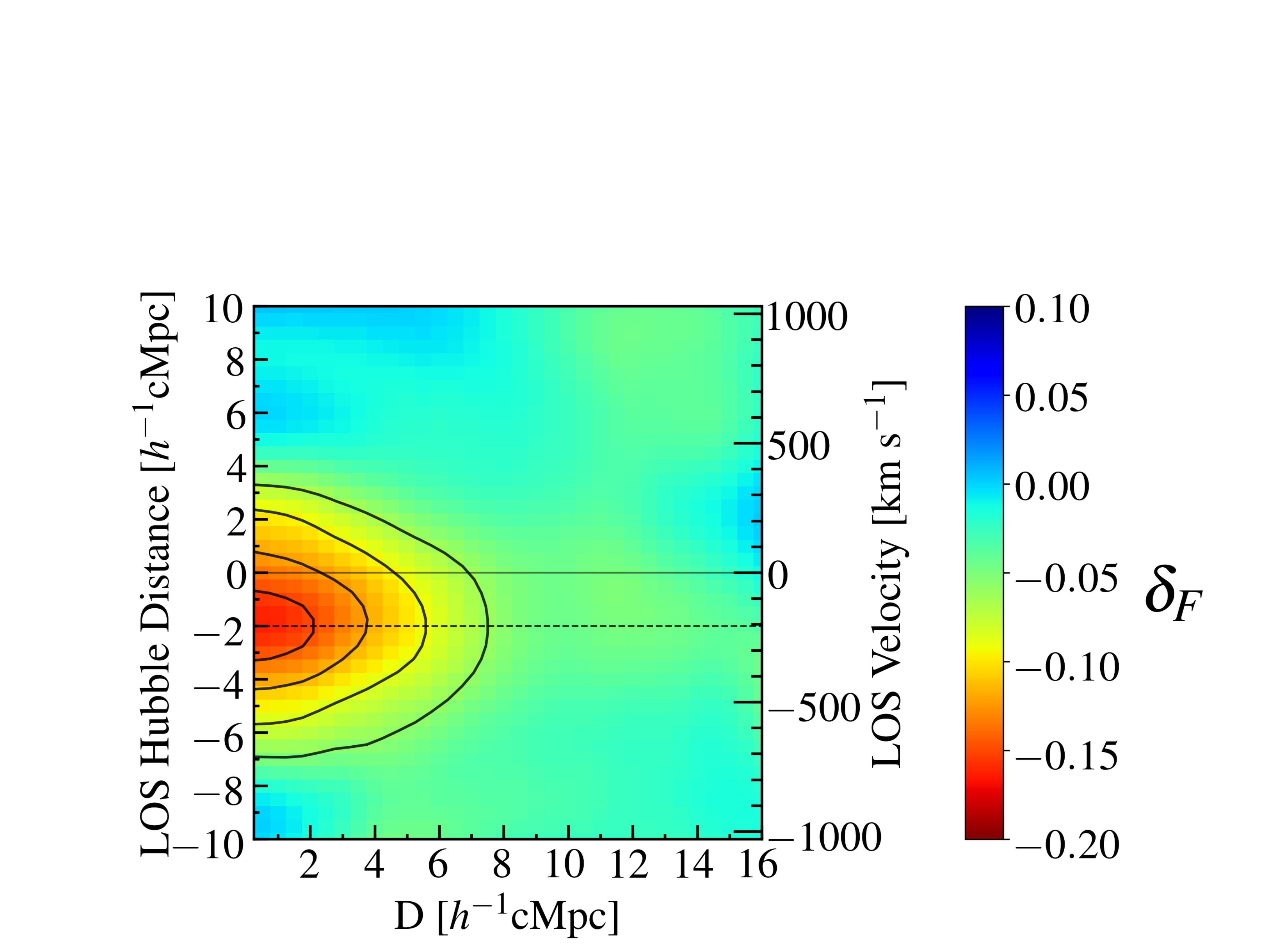}
\caption{
Averaged 2D \hi\ absorption map as a function of transverse $D$ and LOS Hubble distances from our LAEs in the COSMOS field. The bin size is $0.5\ h^{-1} {\rm cMpc}$ and the map is smoothed by a 2D Gaussian kernel with a FWHM of $1.5\ h^{-1} {\rm cMpc}$.
The color code represents the \hi\ transmission overdensity $\delta_{F}$. The black contours indicate the significance levels of the $\delta_{F}$ values (over the errors) from the $2$ to $5$ $\sigma$ levels by a step of $1\sigma$.
The apparent \hi\ transmission overdensity is centered at $-2$ in the LOS Hubble distance (200 km s$^{-1}$ blueshift) due to \lya\ radiative transfer.}
\label{fig: cc2d}
\end{figure*}

Figure \ref{fig: cc2d} shows the 2D \hi\ distribution map of the COSMOS LAEs. 
In Figure \ref{fig: cc2d}, the LAEs are located at $(Z, D) = (0, 0)$ \hmpc. 
There exist \hi\ absorption enhancements around LAEs at $-7\lesssim Z\lesssim 3$ \hmpc\ along the LOS direction and $D\lesssim 8$ \hmpc\ in the transverse direction from the LAEs. 
The \hi\ absorption enhancements have an anisotropic distribution whose \hi\ absorption peak has an offset toward the observer by $\sim 2\ h^{-1} {\rm cMpc}$ corresponding to the blueshift of $\sim200$ km s$^{-1}$ from the LAE Ly$\alpha$ redshifts. This blueshift is consistent with the recent MUSE galaxy-background QSO pair study of \cite{Muzahid2019a}, and is
explained by a well-known velocity offset of a Ly$\alpha$ redshift from a galaxy systemic redshift by $\sim200$ km s$^{-1}$ for LAEs on average
(\citealt{Steidel2010, Hashimoto2013, Shibuya2014b, Song2014}; 
see Ouchi et al. 2020 in press and references therein)
due to \lya\ resonant scattering \cite[e.g.,][]{Neufeld1990a, Dijkstra2006a}.
In other words, our 2D \hi\ distribution map reproduces the average $\sim200$ km s$^{-1}$ offset of Ly$\alpha$ redshifts by the independent analysis that is different from previous studies requiring both galaxy Ly$\alpha$ and systemic redshift determinations from other spectral features.
\par
Figure \ref{fig: cc2d} also shows 
an elongated \hi-gas distribution 
whose absorption enhancements are stronger  
in the transverse direction than along the LOS direction
from the \hi\ absorption peak.
This elongation may originate from large-scale gas infall 
toward the galaxy halos 
as claimed in previous galaxy-background QSO pair studies
\citep[][]{Turner2014a, Bielby2017a} and
as predicted by numerical simulations 
\citep[][]{Turner2017a, Kakichi2018a}.
A detailed analysis with radiative transfer simulations 
for the elongated \hi-gas distribution 
will be presented in a forthcoming publication
(Byrohl et al. in preparation).

\par 
We measure \hrp s around the COSMOS LAEs (Section \ref{sec: lae_catalog}), 
spherically averaging radial profiles of 
\dF\ taken from the COSMOS \hi\ tomography map 
(Section \ref{subsec: tom cosmos}) 
as a function of 3D distance from the LAEs.
This is the similar analysis of \hi-QSO 
performed in a previous study of \citet{Mukae2020a}.
The 3D distances from the LAEs are defined as 
\begin{eqnarray}
R_{\rm 3D} \equiv \sqrt{D^2 + d_{z}^2}, 
\label{eq: 3D distance}
\end{eqnarray}
where $d_{z}$ is the Hubble flow comoving distance
from the LAEs under the assumption that the \hi\ absorbers have zero peculiar velocities relative to the LAEs.  
To estimate uncertainties of the spherically averaged \dF, we use the 1000 mock \hi\ tomography maps (Section \ref{subsec: tom cosmos}). For each mock map, we compute \hrp s averaged over our LAEs, to obtain $68${\%} intervals as $1\sigma$ confidence intervals. 
\par
\begin{figure}[t]
\centering
\includegraphics[width=1.0\hsize,  clip, bb= 0 0 750 650, clip=true]{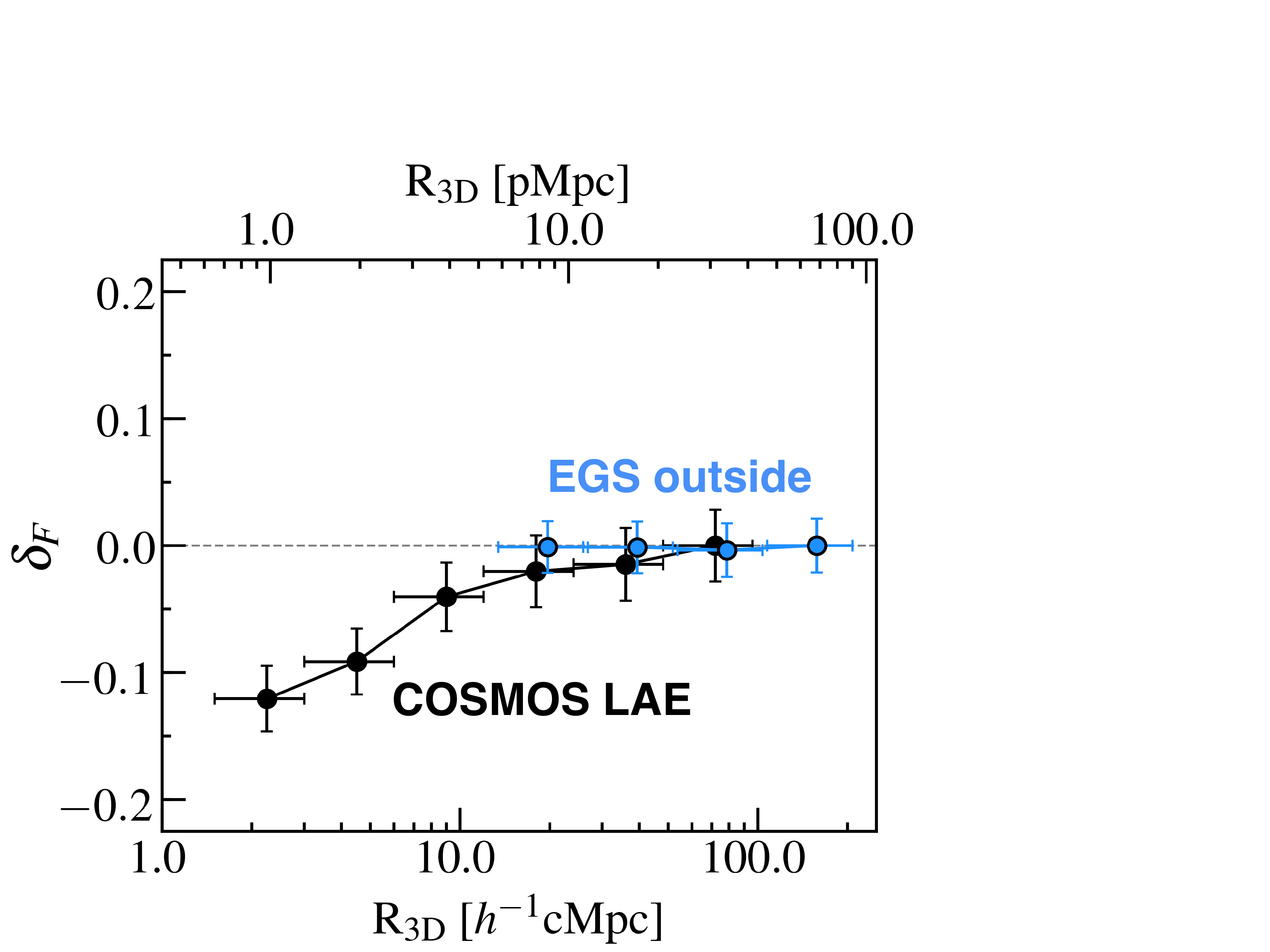}
\caption{
\hrp\ around LAEs in the blank fields.
The black circles are the COSMOS LAEs (Section \ref{sec: results cosmos}).
The blue circles are the EGS LAEs that reside
outside of the QSO overdensity EGS-QO1 
(i.e. EGS outside; Section \ref{sec: results egs}).
The horizontal bars represent the measurement boundaries for the average.
The gray dashed line is the cosmic mean level of the IGM \hi\ absorption at $z=2.3$.
A modest \hi\ transmission overdensity is found near LAEs.}
\label{fig: delf_r_cosmos_lae}
\end{figure}

In Figure \ref{fig: delf_r_cosmos_lae}, the black circles present the average \hrp\ of the COSMOS LAEs. The value of \dF\ decreases (i.e., the \hi\ absorption increases) from the cosmic mean level \dF\ $= 0$ to $-0.1$ with decreasing \rthd\ from $\sim 10$ \hmpc\ to $\sim 1$ \hmpc\ around the LAEs. In other words, our \hrp\ shows strong \hi\ absorption exists around galaxies up to the 10 \hmpc\ scale.
This trend is consistent with the one found for bright LAEs by \cite{Momose2020b},
whose \lya\ luminosity limit for the bright LAEs is $L_{\rm Ly\alpha} \geq 10^{43}$ erg s$^{-1}$, comparable to the \lya\ luminosity limit of the HETDEX LAEs (Section \ref{sec: lae_catalog}).
\par
\begin{figure*}[p]
\begin{tabular}{c}
\begin{minipage}[t]{1.0\hsize}
\centering
\includegraphics[width=1.0\hsize, clip, bb= 0 0 1600 1800, clip=true]{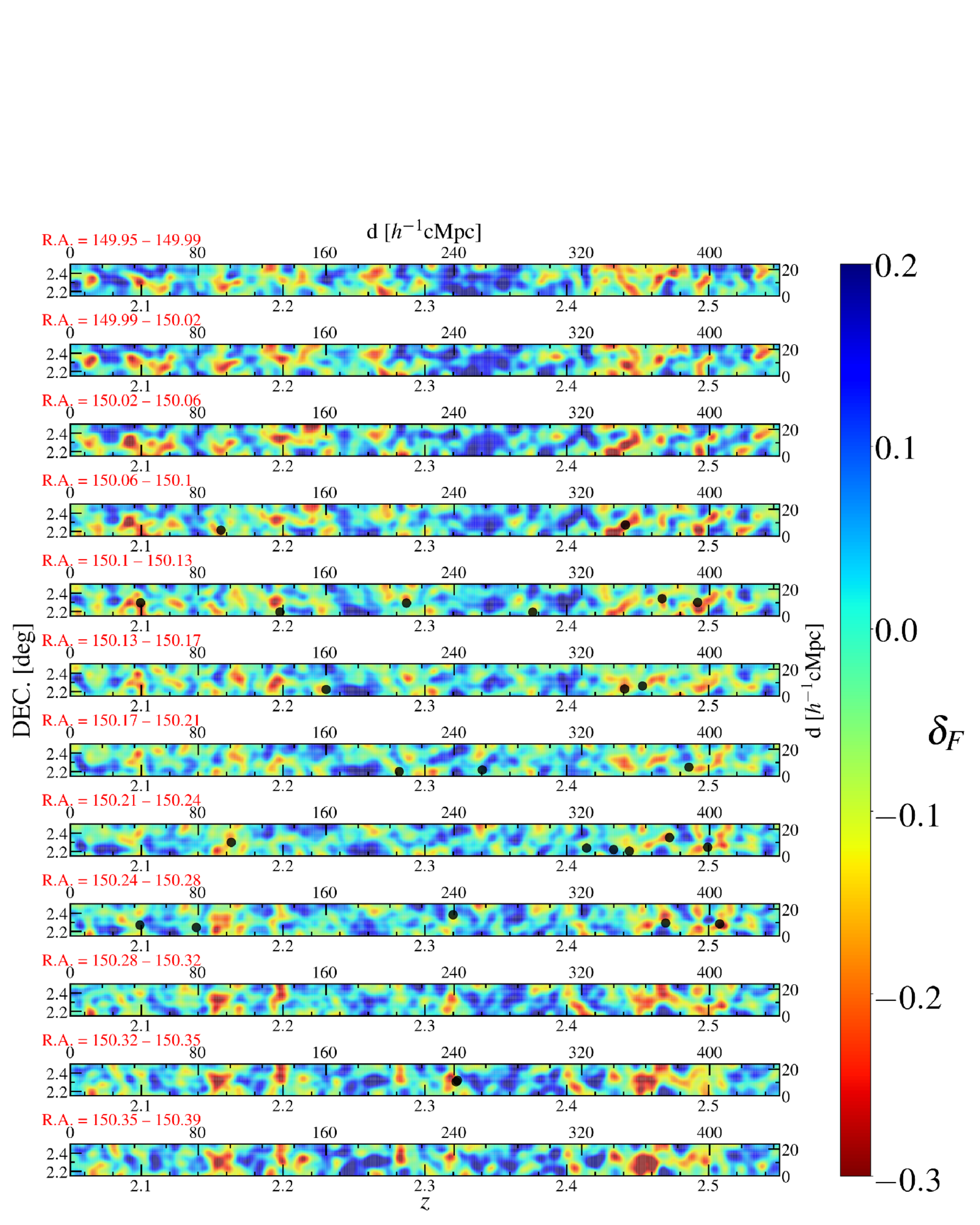}
\caption{
 Projections of $2.5\ h^{-1}$ cMpc-width slices of the \hi\ tomography map in COSMOS over the R.A. direction. The width of the slice is comparable to the spatial resolution of the \hi\ tomography map (Section \ref{subsec: tom cosmos}).
 The color contours represent the \hi\ transmission overdensity $\delta_{F}$: the red (blue) color denotes a negative (positive) $\delta_{F}$ value that corresponds to 
 a strong (weak) \hi\ absorption. The black dots  
 indicate positions of the COSMOS LAEs. Because there are no eBOSS QSOs in this small volume of the COSMOS \hi\ tomography map, no QSOs are shown in this figure.
 Visual inspection may find potential spatial correlations between \hi\ absorption and LAEs, but quantitative analysis is needed to quantify the spatial correlations. }
\label{fig: yz_cosmos}
\end{minipage}\\
\end{tabular}
\end{figure*}

 
All of the results of the two components above indicate \hi\ absorption excesses around galaxies over Mpc scales
(consistent with those of previous galaxy-background QSO pair studies of \citet{Turner2017a}, \citet{Bielby2017a}), 
which confirm the impression of a spatial correlation between \hi\ and galaxies 
seen in Figure \ref{fig: yz_cosmos}.
Note that these results of the spatial correlations are free from influences of bright type-I QSOs, because no eBOSS QSOs at $z=2.05$--$2.55$ are found in the cosmic volume of the COSMOS \hi\ tomography map (Section \ref{sec: foreground_qusasar}).

\subsection{\hi--Gas Distribution Around an Extreme QSO Overdensity
} \label{sec: results egs}
We investigate the IGM \hi-gas distribution around QSOs and galaxies 
in an extreme QSO overdensity.
\par
\begin{figure*}[ht]
\centering
\includegraphics[width=0.9\hsize, clip, bb= 0 0 650 800, clip=true]{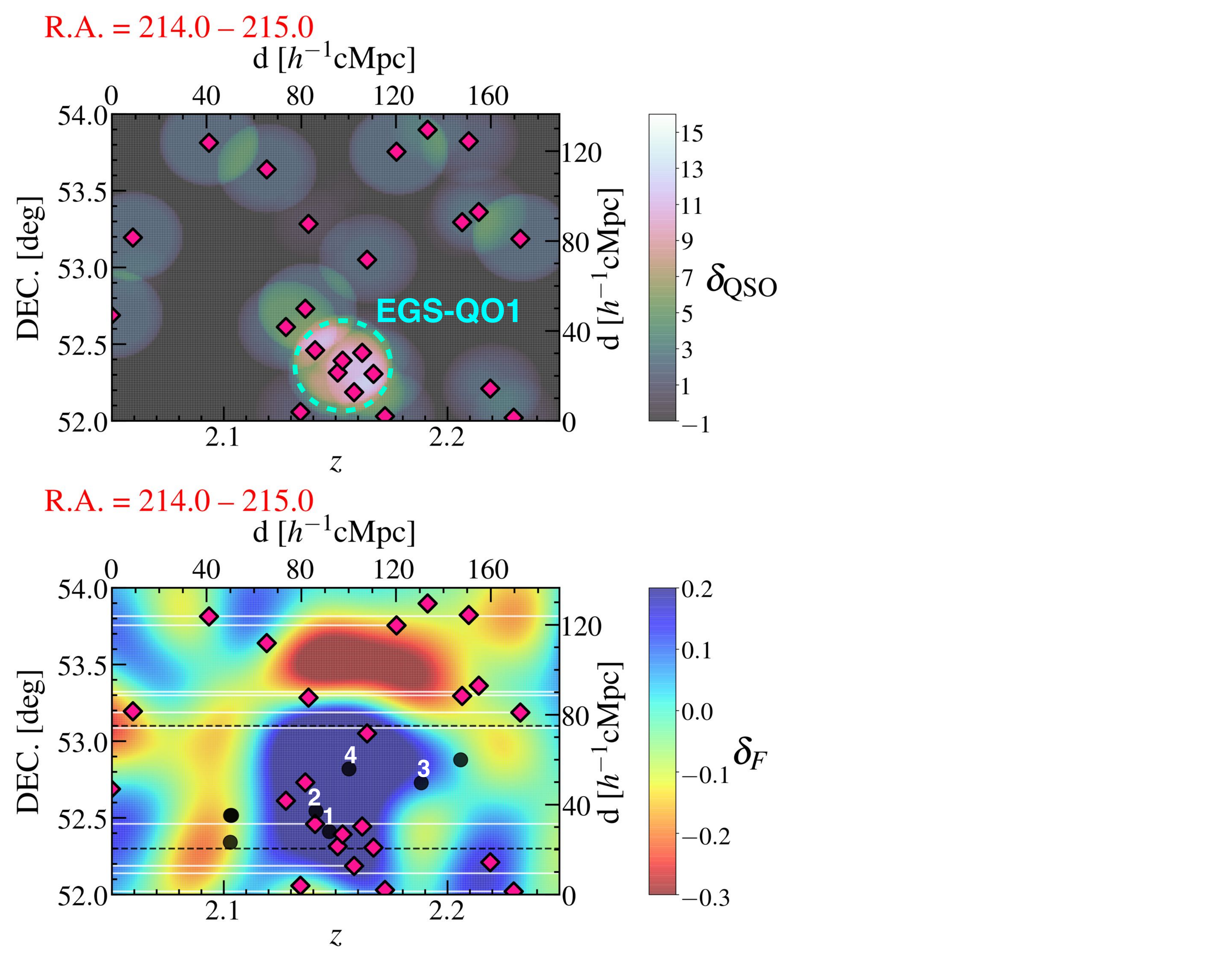}
\caption{
Top panel: 
Projection of the \dQSO\ map for a $40\ h^{-1}$ cMpc (R.A. direction) slice 
in EGS that includes the EGS-QO1 QSO overdensity.
The color contours represent the QSO overdensity \dQSO, 
where white indicates a high \dQSO\ value.
The magenta diamonds represent the foreground QSOs (Section \ref{sec: foreground_qusasar}). 
The dashed circle indicates EGS-QO1, the identified QSO overdensity.  
Bottom panel: Same as the top panel, but for the \hi\ tomography map. 
The width of the slice, $40\ h^{-1}$ cMpc, is twice as large as 
the spatial resolution of this map (Section \ref{subsec: tom egs}). 
The color contours represent the \hi\ transmission overdensity
$\delta_{F}$, where the red (blue) color is 
a negative (positive) $\delta_{F}$ that corresponds to 
a strong (weak) \hi\ absorption.
The white dashed horizontal lines denote the background QSO sightlines.
As in the top panel, the magenta diamonds represent the foreground QSOs
(Section \ref{sec: foreground_qusasar}) residing in the slice. 
The black circles denote the positions of the LAEs (Section \ref{sec: lae_catalog}).
The black dashed lines indicate the edges in declination of 
the LAE detections covered by the HETDEX survey.
Although spatial correlations between objects (QSOs and LAEs) and 
\hi\ absorption may be identified by visual inspection, 
one needs quantitative analysis (Section \ref{sec: results egs}) 
to conclude the reality of the apparent spatial correlations.
}
\label{fig: yz_egs}
\end{figure*}

 
First, we search for QSO overdensities in the large $6.0$-\sqdeg\ area of the EGS field that is sufficient to find rare systems of QSO overdensities \citep[e.g.,][]{cai2017b, Hennawi2015a}. 
We use the 78 foreground QSOs (Section \ref{sec: foreground_qusasar}), and 
estimate QSO overdensities within a sphere of radius 20 \hmpc\ at $z=2.3$. This radius is larger than the one applied for galaxy overdensity measurements \citep{Chiang2013a, Chiang2014a}, because the number density of QSOs is 
about two orders of magnitude smaller than that of galaxies at $z\sim 2$. 
The QSO overdensity is defined as 
\beq 
\delta_{\rm QSO} \equiv \frac{ n_{\rm QSO} }{ \overline{n}_{\rm QSO} } -1, \label{eq:lae overdensity}
\eeq
where $n_{\rm QSO}$ ($\overline{n}_{\rm QSO}$) is the number density (mean number density) of the QSOs in a sphere.
The mean number density is derived in the cosmic volume of 
the \hi\ tomography map. 
The expected number of QSOs in the sphere is about $0.35$ 
comparable to that estimated from QSO luminosity functions \citep[e.g.,][]{Palanque-Delabrouille2013a}
integrated to the detection limit of the eBOSS QSOs ($g\simeq22$ mag).
We calculate \dQSO\ at pixel positions of the EGS \hi\ tomography map, and make a \dQSO\ map whose volume coverage is similar to that of the \hi\ tomography map. Searching for the largest QSO overdensity in the \dQSO\ map, we find an extremely high overdensity at $z=2$, dubbed EGS-QO1, 
whose QSO overdensity is \dQSO\ $=16.2\pm7.0$. 
The top panel of Figure \ref{fig: yz_egs} shows a projection of a $40\ h^{-1}$ cMpc-width slice of our \dQSO\ map with EGS-QO1.
The overdensity of EGS-QO1 is clearly distinguished. 
\par
\begin{figure}[ht]
 \centering
 \includegraphics[width=1.1\hsize, clip, bb= 0 0 600 800, clip=true]{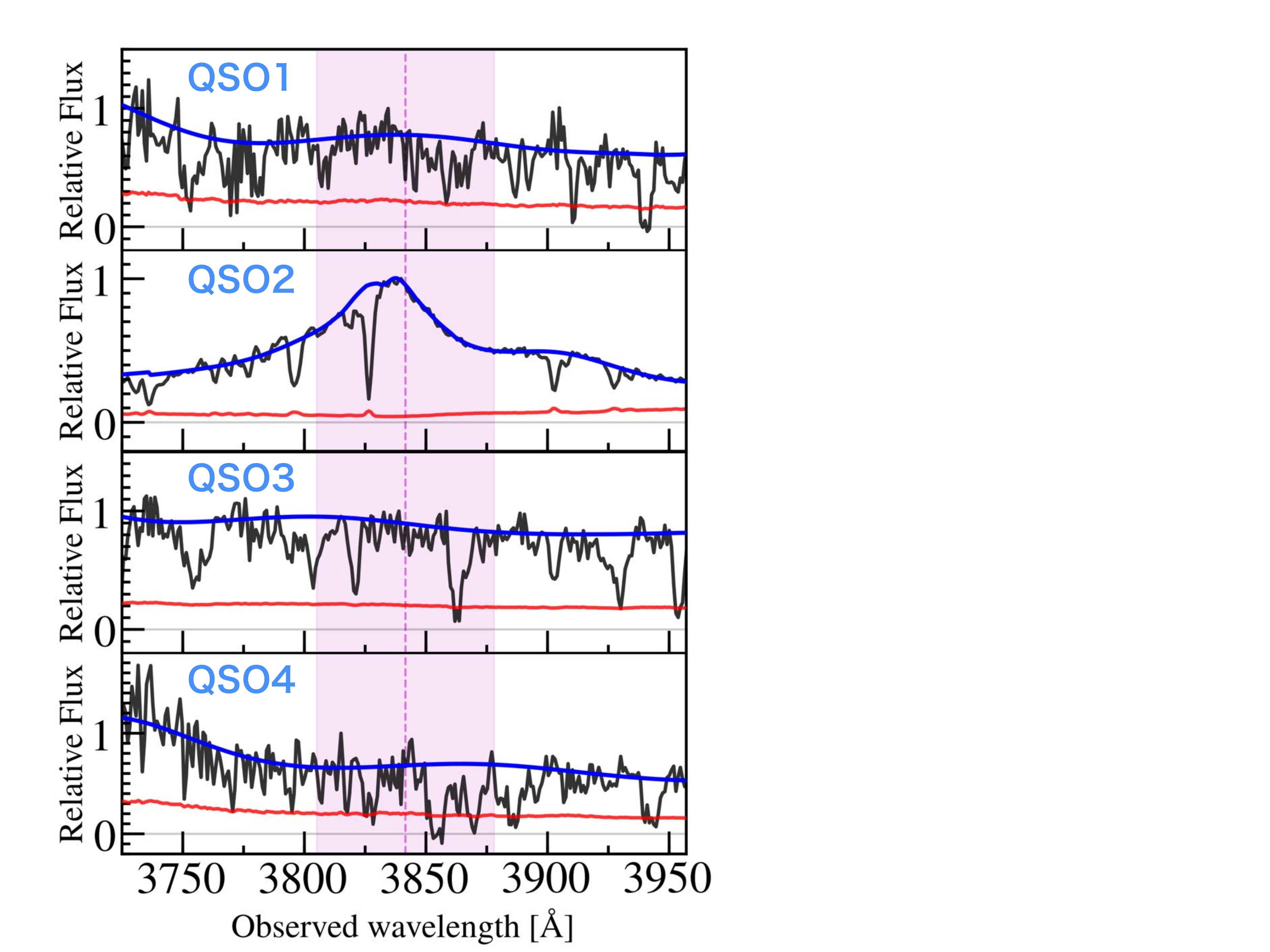}
 \caption{
 Spectra of the background QSO sightlines
 that probe the large \hi\ underdensity around EGS-QO1.
 These background QSO sightlines are labeled QSO1-4 in Figure \ref{fig: sky egs}.
 The black and red lines depict flux and error per pixel, respectively.
 The error is calculated from the errors of the flux measurements and the MF-PCA continuum fitting (Section \ref{subsec: intrinsic continua}).
 The blue curves are the estimated continua.
 The magenta dashed lines (shades) represent the wavelength center (range) that corresponds to the \hi\ underdensity redshift (range) of $z=2.16$ ($z=2.13-2.19$).
 The background QSO sightlines indicate weak \hi\ absorption within $z=2.13-2.19$, 
 although QSO2 is not used in the wavelength range shown in this Figure for the \hi\ tomographic map construction.
 }
 \label{fig: bq spec 2}
\end{figure}

Next, we inspect the \hi\ environment of EGS-QO1 with the \hi\ tomography map. The bottom panel of Figure \ref{fig: yz_egs} shows the projected \hi\ tomography map of the same slice as presented in the top panel of Figure \ref{fig: yz_egs}. 
Comparing the top and bottom panels of Figure \ref{fig: yz_egs}, 
we find that EGS-QO1 resides in an \hi\ underdensity volume with a size of
$\sim 40\ \times 70\ \times 40\ h^{-3} {\rm cMpc}^{3}$ 
at $z=2.13$--$2.19$ (centered at $\left < z \right > = 2.16$).
Red histogram in Figure \ref{fig: hist_delF_egs} depicts 
the \dF\ pixel distribution of EGS-QO1. 
The \hi\ absorption values in EGS-QO1 ranges \dF\ $\simeq 0.1 - 0.4$, 
indicating that the EGS-QO1 resides in an \hi\ underdensity volume.
Note that the typical 1$\sigma$ uncertainty of $\delta_F$ 
for a pixel in the \hi\ tomographic map is $0.1$ (Section \ref{subsec: tom egs}).
Although no background QSOs probe \hi\ absorption at the exact sky center of 
the \hi\ underdensity volume, spectra of several of the QSOs distributed 
over this volume show evidence of the \hi\ underdensity (Figure \ref{fig: bq spec 2}).
The \hi\ underdensity associated with the QSO overdensity suggests that 
the QSOs forming the EGS-QO1 overdensity would make a large ionized bubble, 
where the \hi\ gas is widely photoionized by the strong ionizing radiation 
of the QSOs. 

The ionized bubble may be created by overlap of proximity zones
of the QSOs, each of which should have a typical size of 
$\sim 10-15\ h^{-1}$ cMpc in diameter at $z\sim2$ \citep[e.g.,][]{DOdorico2008a, Mukae2020a, Jalan2019a}.
If the ionized bubble length of $\sim40$ \hmpc\ is 
made by three QSOs roughly distributed 
along the redshift direction in EGS-QO1,  
each QSO would form a proximity zone with $40/3\simeq13$ \hmpc\ in diameter,
which is comparable to typical sizes in the literature.
When the proximity zone size is simply divided by the speed of light,
we obtain QSO lifetimes of $t_{\rm QSO} = 10^{7.3}$ years 
consistent with typical values of $t_{\rm QSO} = 10^{7-9}$ years 
that are constrained by clustering measurements 
\citep[e.g.,][]{Adelberger2005b, White2012a, Conroy2013a}.

Note that the \hi\ absorption is made by the ionized IGM 
with \hi\ fraction as low as $\sim 0.6 \times 10^{-6}$ 
\footnote{The \hi\ fraction is obtained with 
a ratio of the EGS-QO1 value to the cosmic value at $z=2$ \citep{McQuinn2016a}.
The ratio is simply estimated from a ratio of column densities 
that are converted from optical depth of 
\dF\ $\simeq0.2$ (the peak value of the \dF\ fraction in Figure \ref{fig: hist_delF_egs}) and \dF\ $=0.0$ (the cosmic mean value)
under the assumption of optically-thin \lya\ forest clouds 
\citep{Draine2011a, Mo2010a}.}.
Even outside of the ionized bubble, 
the IGM is highly ionized with the cosmic average \hi\ fraction of 
$\sim  2.0 \times 10^{-6}$ at $z=2$ \citep{McQuinn2016a}.
The ionized bubble defined here is the cosmic volume 
with small \hi\ fraction 
that is probably caused by the strong QSO radiation.
\par
\begin{figure}[ht]
\centering
\includegraphics[width=1.0\hsize,  clip, bb= 0 0 750 650, clip=true]{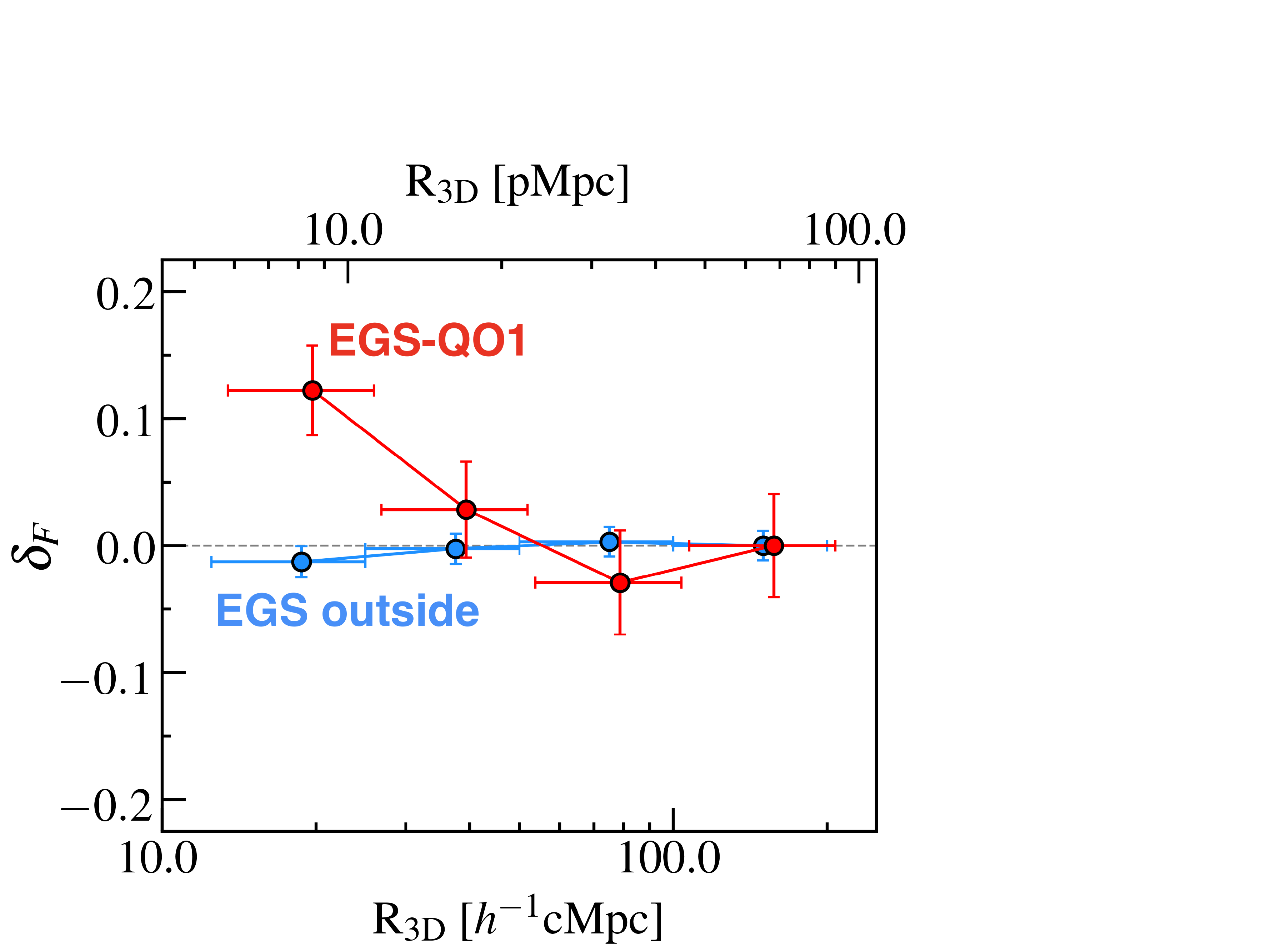}
\caption{
Same as Figure \ref{fig: delf_r_cosmos_lae}, but for the six QSOs forming the extreme QSO overdensity, EGS-QO1 (red circles).
We overplot the \hrp\ of the 72 foreground QSOs that reside outside of the EGS-QO1 (i.e. EGS outside; blue circles).
A large-scale \hi\ transmission underdensity is found around the QSOs of EGS-QO1.
It is noted that the EGS \hi\ tomography map has a smoothing scale of 20 \hmpc\ and 
the \hi\ radial profiles are constructed beyond 20 \hmpc\ diameter around QSOs.
}
\label{fig: delf_r_egs_qso}
\end{figure}

We then measure the \hrp\ averaged over the six QSOs forming the EGS-QO1 overdensity. Figure \ref{fig: delf_r_egs_qso} represents the average \hrp\ around the six QSOs. 
The \dF\ values increase (i.e., \hi\ absorption weakens) from the cosmic mean level (\dF $=0$) to \dF $=0.1$
with decreasing \rthd\ from $\sim 100$ \hmpc\ to $\sim 10$ \hmpc\ around the QSOs.
This \hrp\ suggests strongly-suppressed \hi\ absorption around the QSOs in the extreme QSO overdensity. It is a clear contrast with the results of the blank fields; the \hrp\ of the rest of 72 QSOs that reside outside of the EGS-QO1 (here after referred to as EGS outside; Figure \ref{fig: delf_r_egs_qso})
and the \hrp\ of galaxies in the COSMOS field (Figure \ref{fig: delf_r_cosmos_lae}). 
\par
\begin{figure*}[p]
\centering
\includegraphics[width=1.0\hsize, clip, bb= 0 0 1200 1200, clip=true]{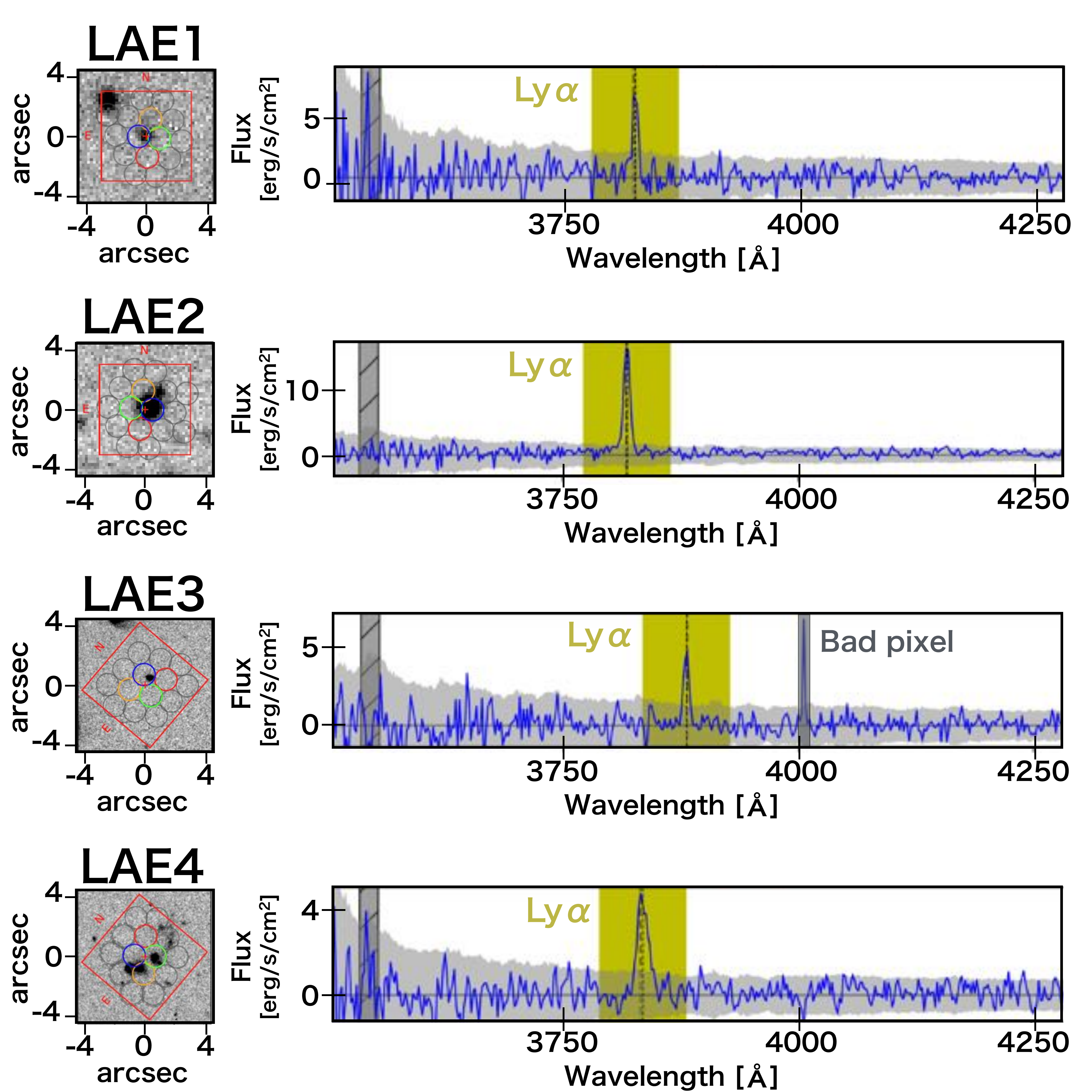}
\caption{
Right: HETDEX spectra of LAE1--4
residing in the extreme QSO overdensity EGS-QO1. 
The black dashed lines denote the wavelengths of the \lya\ emission lines that are highlighted with the yellow shades. The light gray shades represent the 3 sigma noise levels. The dark gray shades present the wavelength range of bright sky lines and bad pixels.
Left: HETDEX Fiber positions atop the broadband images taken with CFHT/MegaCam or HST/WFC3. The red, blue, green, and yellow circles present the positions and the sky coverage of the fibers used for the measurements of the LAE spectra.}
\label{fig: lae_egs}
\end{figure*}
\begin{figure}[ht]
\centering
\includegraphics[width=1.0\hsize,  clip, bb= 0 0 750 650, clip=true]{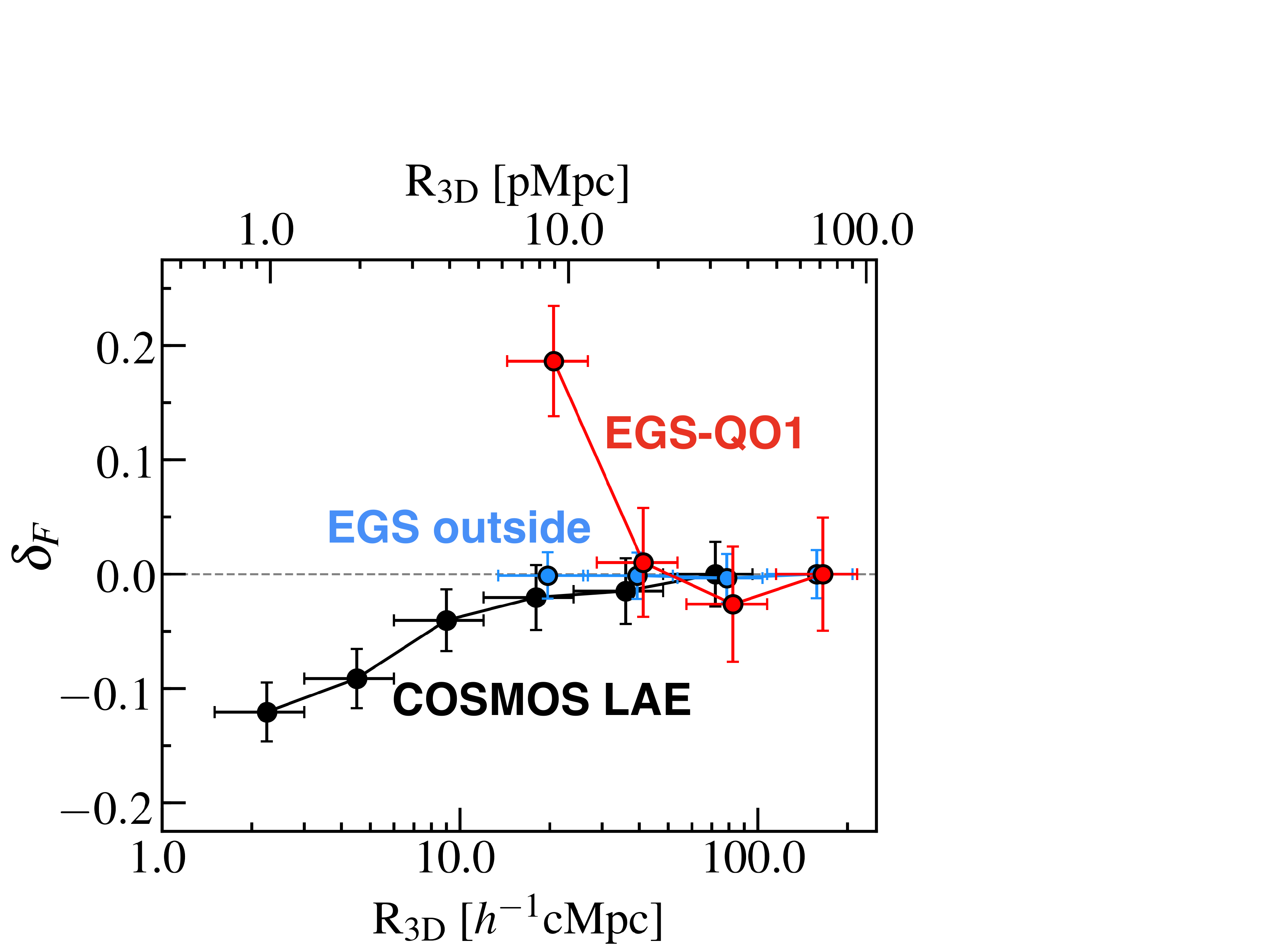}
\caption{
Same as Figure \ref{fig: delf_r_cosmos_lae}, 
but for the \hrp\ around LAE1--4 
in the the extreme QSO overdensity EGS-QO1 (red circles).
The blue and black circles are the \hrp s of LAEs in the blank fields: 
the blue circles are the rest of 22 EGS LAEs that reside outside of the EGS-QO1 (i.e. EGS outside) and the black circles are the COSMOS LAEs.
A significant \hi\ transmission underdensity is identified around LAE1--4, which is similar to the one around the QSOs in EGS-QO1 
(Figure \ref{fig: delf_r_egs_qso}).}
\label{fig: delf_r_egs_lae}
\end{figure}

Here we need to examine whether this contrast is made by the choices of 
the \hrp\ measurement centers (QSOs vs. galaxies) or 
the environment (QSO overdensity vs. blank field).
In the cosmic volume of the EGS \hi\ tomography map, we find
26 LAEs, four of which reside in the EGS-QO1 QSO overdensity
at $z=2.13$--$2.19$. These four LAEs, dubbed LAE1-4 are shown 
with black circles and white labels in Figure \ref{fig: yz_egs}. 
In EGS-QO1, the density of LAEs is high, and there is a moderate LAE 
overdensity of $\sim 1$ in a radius of 20 \hmpc. Figure \ref{fig: lae_egs}
presents the HETDEX spectra and CFHT/HST images of LAE1-4. 
We calculate \hrp s around LAE1-4 in the EGS-QO1 QSO overdensity 
in the same manner as Section \ref{sec: results cosmos}, and show 
the average \hrp\ in Figure \ref{fig: delf_r_egs_lae}. We find that this \hi\ radial profile around the LAEs (Figure \ref{fig: delf_r_egs_lae}) is similar to that of the QSOs in EGS-QO1 (Figure \ref{fig: delf_r_egs_qso}) and clearly different from the one around LAEs in the blank field (Figure \ref{fig: delf_r_cosmos_lae}).
We also measure the \hrp s around the rest of 22 LAEs that reside 
outside of EGS-QO1 (i.e. EGS outside; Figure \ref{fig: delf_r_cosmos_lae}), and find that the \hi\ radial profile is consistent with the one in the blank field on the scale down to $\sim 20$ \hmpc\, the resolution limit of the EGS \hi\ tomography map.
In other words, the \hi\ absorption is significantly weakened around galaxies in the QSO overdensity EGS-QO1, in contrast with the strong \hi\ absorption around galaxies in the blank field (Section \ref{sec: results cosmos}). The weak \hi\ absorption around galaxies found in EGS-QO1 is probably caused by the environment of the QSO overdensity that produces the ionized bubble.

\par
\begin{figure*}[p]
\centering
\includegraphics[width=1.0\hsize, clip, bb= 0 0 1000 1000, clip=true]{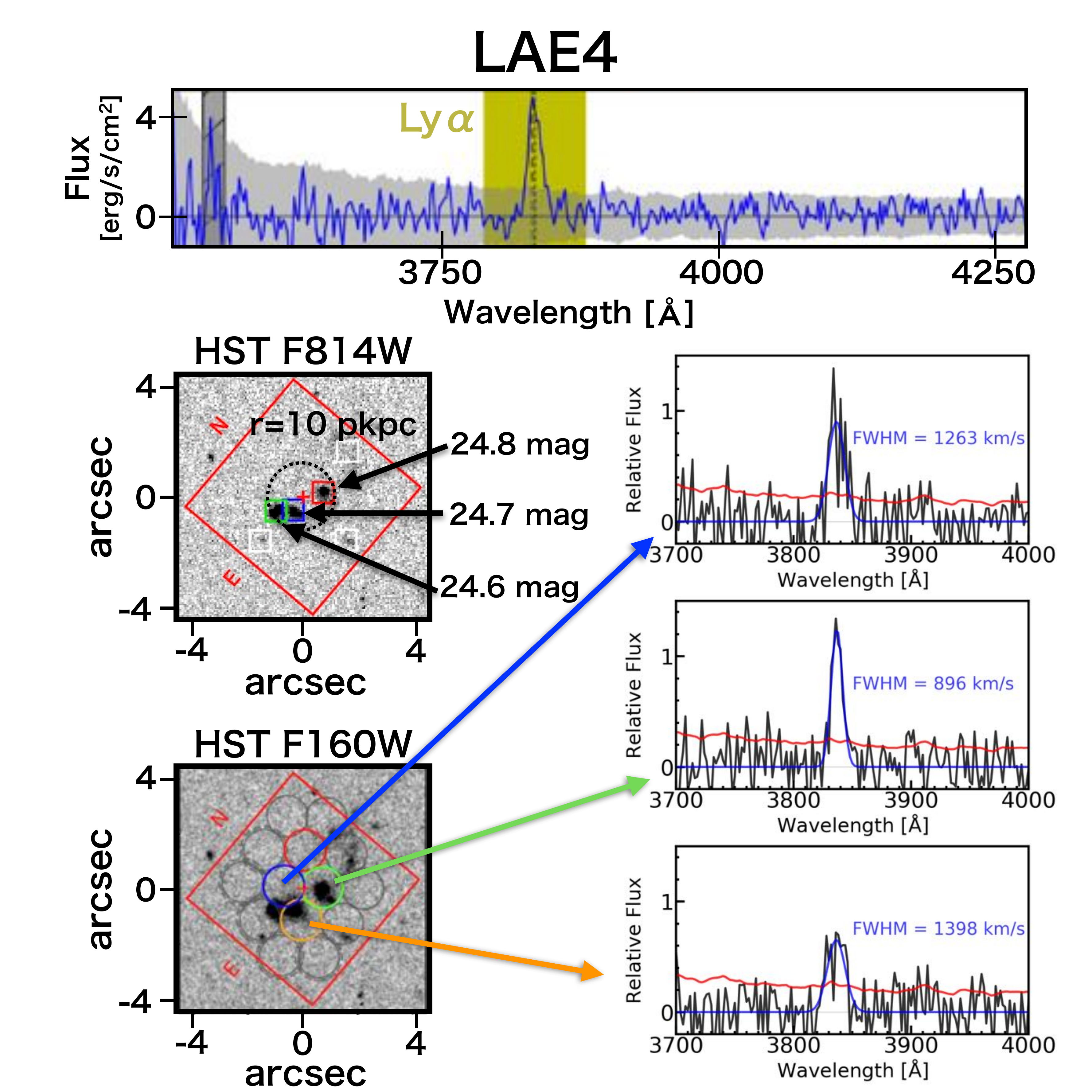}
\caption{
Top: the spectrum of the LAE4 over the HETDEX fibers. 
The black dashed line represents the wavelength of the \lya\ emission line that is highlighted with the yellow shading.
Middle: HST F814W image of LAE4. The red, blue, and green squares indicate positions of the triple-continuum components, all of which have similar F814W magnitudes, $24.8$, $24.7$, and $24.6$ mag, respectively. The dashed circle (red cross) denotes the 10 pkpc circle (circle center) indicating the approximate size of the LAE4 system.
Bottom: Positions and sky coverage of the HETDEX fibers, shown as circles on the HST F160W image. The spectra associated with the blue, green, and orange fibers have clear detections of \lya\ emission lines that are presented in the spectrum panels on the right-hand side. In these spectrum panels, the black, red, and blue lines denote the spectra, the $1\sigma$ sky levels, and the best-fit Gaussian profiles to the \lya\ emission lines, respectively. 
Note that the 1.5 arcsecond diameter fibers and the image quality of the HETDEX observation are much larger than the image sampling in the HST images, so the spectra of the three HST continuum image components are blended.}
\label{fig: agntriplet}
\end{figure*}

 
We find that one of the four LAEs, LAE4, shows triple-continuum components over the scale of a 10 pkpc-radius circle on the HST F814W/F160W images of the EGS, indicative of a triple merging system 
(middle/bottom panels of Figure \ref{fig: agntriplet}). 
LAE4 is located at (R.A., Decl.)$=$(14:19:13.02, +52:49:11.2) near the center of the ionized bubble of EGS-QO1 (bottom panel of Figure \ref{fig: yz_egs}). 
The average \lya\ redshift of LAE4 over the HETDEX fibers is 
$z_{\rm Ly\alpha}=2.156$, and the total \lya\ luminosity over the fibers is
$L_{\rm Ly\alpha}^{*} = 10^{43.06}$ erg s$^{-1}$. 
The top panel of Figure \ref{fig: agntriplet} presents 
the spectrum of LAE4, summed over the fibers,
which is the same as the one in Figure \ref{fig: lae_egs}. The spectrum of LAE4 has a broad \lya\ emission line whose 
FWHM is $\sim 1100$ km s$^{-1}$
(about 3 times broader than the instrumental resolution), 
possibly suggesting a type-I AGN 
(e.g. Kakuma et al. in preparation)\footnote{Other AGN features of high-ionization emission lines, 
C{\sc iv} $\lambda1549$ \AA\ and He{\sc ii} $\lambda1640$ \AA,
are not detected in the LAE4 spectrum shown in Figure \ref{fig: lae_egs}.
This is probably because the LAE4 is optically faint ($\sim$ 25 mag) 
and its metal lines are too faint to be identified.}. 
The bottom panel of Figure \ref{fig: agntriplet} presents the positions and
the sky coverage of the fibers used for the measurements of
the LAE4 \lya\ redshift and total luminosity, 
and indicates three fibers cover the triple-continuum components with the blue, green, and yellow circles. The spectra of these three fibers are shown
on the right-hand side in Figure \ref{fig: agntriplet}.
All of the three fiber spectra of the blue, green, and yellow circles have
\lya\ emission lines at $z_{\rm Ly\alpha}=2.16$, suggesting that 
the three objects probably reside at the same redshift. 
Moreover, these three fiber spectra have broad \lya\ emission lines with
an FWHM of $\sim 1000$ km s$^{-1}$. 
Although each of these fibers does not separately cover one of the triple-continuum components
whose blending due to the HET image quality and large fiber diameter may produce apparent broad lines,
these broad \lya\ emission lines found in the different positions over the triple-continuum components imply that the LAE4 system could be a merging type-I AGN doublet or triplet.

\section{Discussion} \label{sec: discussion}
\begin{figure*}[p]
\centering
\includegraphics[width=1.0\hsize, clip, bb= 100 600 1500 2000, clip=true]{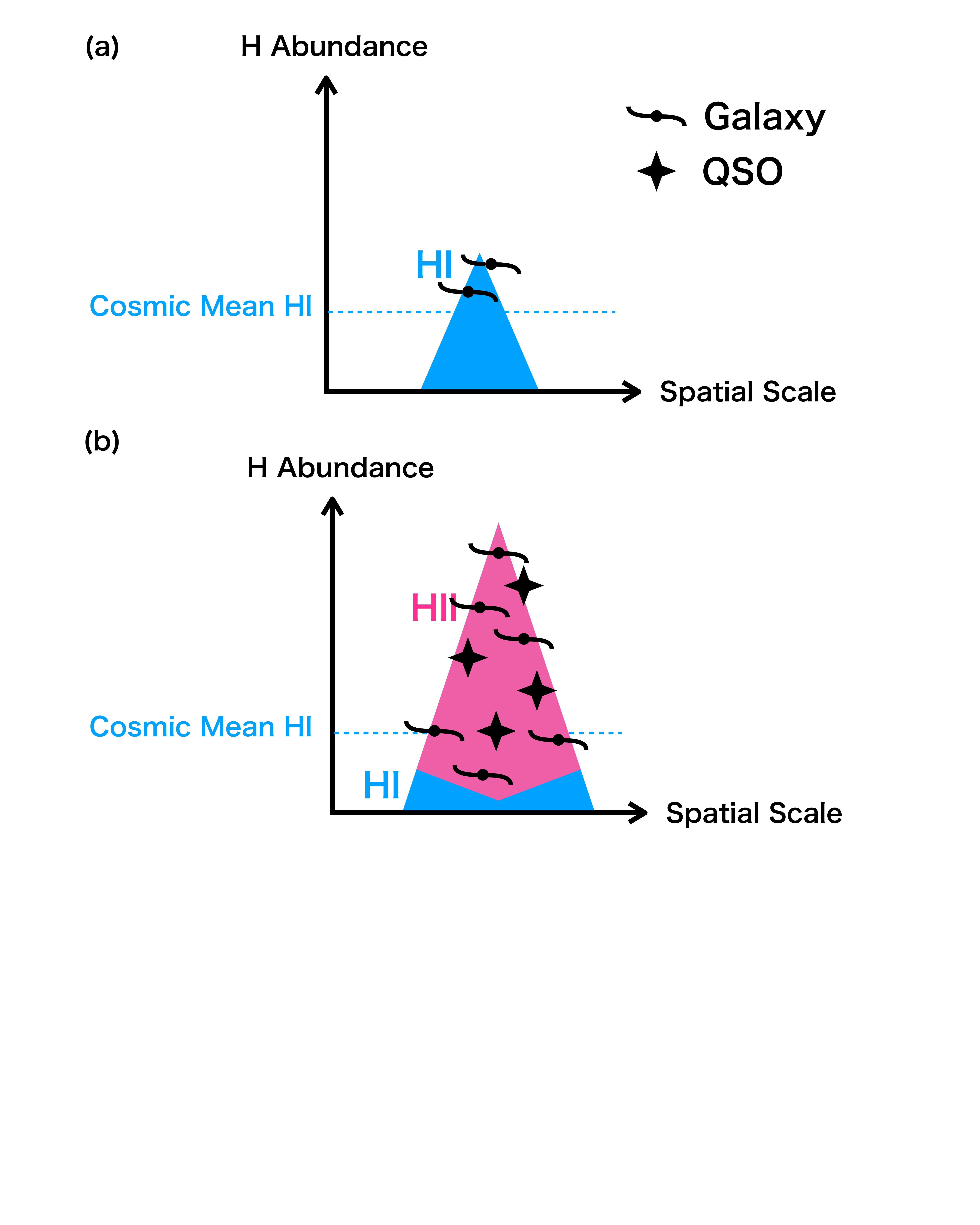}
\caption{
Schematic illustration of the physical picture suggested by this study. The images (a) and (b) show the ionized and neutral hydrogen gas around galaxies/QSOs with red and blue colors, respectively; (a) represents galaxies in an average blank field, while (b) indicates galaxies/QSOs in an extreme QSO overdensity.
}
\label{fig: schematic picture}
\end{figure*}

In Section \ref{sec: results cosmos}, 
we have investigated spatial correlations between IGM \hi-gas and galaxies in the COSMOS field, the blank field with no QSO overdensities.
We have found that strong \hi\ absorption exists around galaxies up to the 10 \hmpc\ scale. 
The result suggests a picture where a galaxy resides in 
an \hi-gas overdensity in a blank field 
as illustrated in Figure \ref{fig: schematic picture} (a).
\par 
In Section \ref{sec: results egs}, 
we have studied spatial correlations between IGM \hi-gas and galaxies 
in the EGS field, where the extreme QSO overdensity EGS-QO1 of six QSOs is found.
There is also a galaxy overdensity in EGS-QO1, traced by LAEs. 
In the \hi\ tomography map of the EGS field, 
EGS-QO1 resides in an ionized bubble 
with a size of $\sim 40$ \hmpc\ at $\left < z \right > = 2.16$.
In fact, the galaxies and the QSOs in and around EGS-QO1 
statistically show weak \hi\ absorption. 
Because \hi\ absorption is weakened in the EGS-QO1 QSO overdensity 
unlike in the blank field, we infer that the QSOs of EGS-QO1 
probably produce an ionized bubble
as a result of the overlap of multiple proximity zones of the QSOs, 
where a galaxy overdensity with a large 
\hi-gas overdensity originally existed.
This physical picture is illustrated in Figure \ref{fig: schematic picture} (b).
\par
The relationship between the physical pictures of 
Figure \ref{fig: schematic picture} (a) and (b) 
may be explained by evolution of photoionization of \hi\ gaseous LSSs.
A matter overdensity in LSSs produces the overdensities of 
\hi\ gas and galaxies \citep{cai2017b, Cai2016a}. 
Once QSO activity is triggered in the galaxies, 
\hi\ gas in the overdensity is photoionized 
by photons from the QSOs \citep{Mukae2020a, Momose2020b}. 
\footnote{
A timeline of this evolutional process is not clear. It may be possible for a large \hi-gas overdensity (as in Figure \ref{fig: schematic picture} b) to form first almost completely, and that subsequently QSOs appear to 
ionize the \hi\ gas. However, it is more likely that QSOs gradually ionize 
the \hi\ gas at the assembly stage of the large \hi-gas overdensity. }
This makes a cosmic volume of weak \hi\ absorption corresponding to the ionized bubble.
Note that again the ionized bubble has a low \hi\ fraction that can be even smaller than that of the cosmic average \hi\ fraction at $z\sim 2$ in the universe after cosmic reionization (Section \ref{sec: results egs}).

\section{Summary} \label{sec: summary}
We have investigated IGM \hi\ gas distributions around $z\sim2$ galaxies 
in two galaxy environments: 
a blank field (COSMOS) and an extreme QSO overdensity field (EGS). 
Combining the large survey datasets of 
galaxies and QSOs that are provided by HETDEX LAEs and SDSS-IV/eBOSS QSOs, respectively,
we construct the samples of foreground galaxies and QSOs at $z\sim2$ as well as the sample of background QSOs at $z>2$. 
In the sample of foreground QSOs, we have identified the extreme QSO overdensity, EGS-QO1, consisting of six QSOs in a radius of 20 \hmpc.
\par
For the COSMOS field, we use the \hi\ tomography map of \cite{Lee2018a}, while for the EGS field we reconstruct 3D \hi\ LSSs by performing \hi\ tomography based on \hi\ absorption found in the spectra of background QSOs.
These COSMOS and EGS \hi\ tomography maps have cosmic volumes of ($30 \times 24 \times 444$, $124 \times 136 \times 444$) $\ h^{-3}$ cMpc$^3$, respectively, at $z=2.3$. 
We have investigated spatial correlations between the \hi\ absorption and the galaxies in the two \hi\ tomography maps. 
Our findings are listed below.

\begin{enumerate}
\item 
In the blank field of COSMOS 
that has no eBOSS QSOs within the volume of the tomography map, 
the spherically averaged \hi\ radial profiles 
indicate that \hi\ absorption around galaxies 
is stronger than those of the cosmic average 
at a distance from these galaxies 
up to $10$ \hmpc. 
Stronger \hi\ absorption is found closer to the galaxies. 
The same trends are also found in the averaged 2D \hi\ absorption map 
(transverse vs. LOS distances). 
These results suggest that the IGM \hi\ gas and galaxies (LAEs) are 
spatially correlated, and that more \hi\ gas exists around such galaxies.
\item 
In the averaged 2D \hi\ absorption map shown in Figure \ref{fig: cc2d}, 
there is an anisotropy in the transverse and LOS directions. On average, 
the \hi\ absorption peak is blueshifted by $\sim 200$ km s$^{-1}$ from the galaxy \lya\ redshift. This result independently reproduces the known average velocity offset between the \lya\ emission redshift and the galaxy systemic redshift, using a completely independent tracer.
\item 
The extreme QSO overdensity of EGS-QO1 resides in an \hi\ underdensity volume 
with a size of $\sim 40\ \times 70\ \times 40\ h^{-3} {\rm cMpc}^{3}$ 
at $z=2.13$--$2.19$ (centered at $\left < z \right > = 2.16$). 
In this volume, the spherically-averaged \hi\ radial profiles show that 
\hi\ absorption around galaxies (and QSOs) is weaker than that of the cosmic average, and that the weaker \hi\ absorption exists closer to the galaxies (and the QSOs).
These results contrast 
with those of the blank fields of COSMOS and EGS outside of EGS-QO1.
Interestingly, in the EGS \hi\ tomography map, 
we identify an ionized bubble with a size of $\sim 40$ \hmpc\ 
at $\left < z \right > = 2.16$ in the volume of the EGS-QO1.
The ionized bubble may form due to intense ionizing photon radiation 
as a result of the overlap of multiple proximity zones of the QSOs.
\item
As noted above, we find possible opposite trends of the \hi-galaxy spatial correlation in the two fields, the blank field and the extreme QSO overdensity field. 
A schematic illustration of our interpretation is shown in Figure \ref{fig: schematic picture}.
Although matter overdensities produce galaxies and galaxy+QSO overdensities, QSOs, if present, ionize the hydrogen gas around galaxies in the overdensity. In an extreme QSO overdensity, the negative correlation of the  \hi -galaxy spatial distribution is probably created by ionizing radiation of the QSOs. If our interpretation (Figure \ref{fig: schematic picture}) is correct, the two different trends of \hi -galaxy spatial correlation may be explained by evolution of photoionization in \hi\ gaseous LSSs. In other words, once QSO activity emerges in galaxies residing in an \hi\ gas overdensity, the \hi\ gas around the galaxies is photoionized by the ionizing photons of the QSOs. 
\end{enumerate}

The evolutionary picture is based on one QSO overdensity (EGS-QO1).  
More QSO overdensities should be investigated to statistically test the picture 
as well as the morphology of giant ionized bubbles, 
because QSOs are actually complicated systems 
whose proximity zones relate to physical quantities 
such as the number of ionizing photons, the opening angle for ionizing photon escape, the lifetime, and the duty cycle \citep[e.g.,][]{Bosman2020a, Adelberger2004a}.
Further investigation will be made with forthcoming data from the HETDEX survey.
Future data releases are expected to have improved spectral traces, sky subtraction, cosmic-ray removal, and flux calibration
(Gebhardt et al. 2020, in preparation).
The HETDEX survey will ultimately provide $10^6$ galaxies 
with spectroscopic redshifts at $z=1.9$--$3.5$ in a $9$ Gpc$^3$ volume.
The HETDEX survey will statistically reveal the galaxy - IGM \hi\ relation 
as a function of QSO overdensities in the large HETDEX Spring Field 
(300 deg$^2$; \citealt{Hill2016a}).
Such statistical studies of LSS photoionization will shed light not only on the suppression of the formation of low-mass galaxies due to enhanced UVB radiation, 
but also the ionization processes of the intra-cluster media of galaxy clusters.

\acknowledgments
We thank the anonymous referee for constructive comments and suggestions that definitely improved the clarity of the paper. 
We are grateful to Koki Kakiichi, Siddhartha Gurung-Lopez, and Chris Sneden 
for their useful discussions and comments. 
\par
HETDEX is led by the University of Texas at Austin McDonald Observatory and Department of Astronomy with participation from the Ludwig-Maximilians- Universit\"{a}t M\"{u}nchen, Max-Planck-Institut f\"{u}r Extraterrestrische-Physik (MPE), Leibniz-Institut f\"{u}r Astrophysik Potsdam (AIP), Texas A\&M University, Pennsylvania State University, Institut f\"{u}r Astrophysik G\"{o}ttingen, The University of Oxford, Max-Planck-Institut f\"{u}r Astrophysik (MPA), The University of Tokyo and Missouri University of Science and Technology. In addition to Institutional support, HETDEX is funded by the National Science Foundation (grant AST-0926815), the State of Texas, the US Air Force (AFRL FA9451-04-2- 0355), and generous support from private individuals and foundations.
The observations were obtained with the Hobby-Eberly Telescope (HET), which is a joint project of the University of Texas at Austin, the Pennsylvania State University, Ludwig-Maximilians-Universit\"{a}t M\"{u}nchen, and Georg-August-Universit\"{a}t G\"{0}ttingen. The HET is named in honor of its principal benefactors, William P. Hobby and Robert E. Eberly.
VIRUS is a joint project of the University of Texas at Austin, Leibniz-Institut f\"{u}r Astrophysik Potsdam (AIP), Texas A\&M University, Max-Planck-Institut f\"{u}r Extraterrestrische-Physik (MPE), Ludwig-Maximilians- Universit\"{a}t M\"{u}nchen, The University of Oxford, Pennsylvania State University, Institut f\"{u}r Astrophysik G\"{o}ttingen,,  Max-Planck-Institut f\"{u}r Astrophysik (MPA)
The Texas Advanced Computing Center (TACC) at The University of Texas at Austin provided high performance computing, visualization, and storage resources that contributed to the research results reported within this paper. 
\par
The Institute for Gravitation and the Cosmos is supported by the Eberly College of
Science and the Office of the Senior Vice President for Research at The
Pennsylvania State University.
\par
This paper is supported by World Premier International Research Center Initiative (WPI Initiative), MEXT, Japan, and KAKENHI (15H02064, 17H01110, and 17H01114) 
Grant-in-Aid for Scientific Research (A) 
through Japan Society for the Promotion of Science.
S.M acknowledges support from the JSPS through the JSPS Research Fellowship for Young Scientists.
\par
Funding for the Sloan Digital Sky Survey IV has been provided by the Alfred P. Sloan Foundation, the U.S. Department of Energy Office of Science, and the Participating Institutions. SDSS-IV acknowledges
support and resources from the Center for High-Performance Computing at
the University of Utah. The SDSS web site is www.sdss.org.
SDSS-IV is managed by the Astrophysical Research Consortium for the 
Participating Institutions of the SDSS Collaboration including the 
Brazilian Participation Group, the Carnegie Institution for Science, 
Carnegie Mellon University, the Chilean Participation Group, the French Participation Group, Harvard-Smithsonian Center for Astrophysics, 
Instituto de Astrof\'isica de Canarias, The Johns Hopkins University, Kavli Institute for the Physics and Mathematics of the Universe (IPMU) / 
University of Tokyo, the Korean Participation Group, Lawrence Berkeley National Laboratory, 
Leibniz Institut f\"ur Astrophysik Potsdam (AIP),  
Max-Planck-Institut f\"ur Astronomie (MPIA Heidelberg), 
Max-Planck-Institut f\"ur Astrophysik (MPA Garching), 
Max-Planck-Institut f\"ur Extraterrestrische Physik (MPE), 
National Astronomical Observatories of China, New Mexico State University, 
New York University, University of Notre Dame, 
Observat\'ario Nacional / MCTI, The Ohio State University, 
Pennsylvania State University, Shanghai Astronomical Observatory, 
United Kingdom Participation Group,
Universidad Nacional Aut\'onoma de M\'exico, University of Arizona, 
University of Colorado Boulder, University of Oxford, University of Portsmouth, 
University of Utah, University of Virginia, University of Washington, University of Wisconsin, Vanderbilt University, and Yale University.

\bibliography{bibliography.bib}

\end{document}